\documentclass[prd,letterpaper]{revtex4}%ws-ijmpe

\usepackage{graphicx}
\usepackage{dcolumn}
\usepackage{amsmath,amssymb,amsfonts}
\usepackage{bm}
\usepackage{epsfig}

\begin{document}

\markboth{Vincent Mathieu,Vicente Vento and Nikolai Kochelev}{Glueballs}

%\catchline{}{}{}{}{}
%%%%%%%%%%%%%%%%%%%%%%%%%%%%%%%%%%%%%%%%%%%%%%%%%%%%%%%%%%%%%%%%%%%%

\title{The Physics of Glueballs}
\author{Vincent Mathieu}
\address{Groupe de Physique Nucl\'{e}aire Th\'{e}orique, Universit\'{e} de Mons-Hainaut,
Acad\'{e}mie universitaire Wallonie-Bruxelles, Place du Parc
20, BE-7000 Mons, Belgium.\\
vincent.mathieu@umh.ac.be}
\author{Nikolai Kochelev}
\address{
Bogoliubov Laboratory of Theoretical Physics,
Joint Institute for Nuclear Research, Dubna, Moscow region, 141980
Russia.\\
kochelev@theor.jinr.ru}
\author{Vicente Vento}
\address{
Departament de F\'{\i}sica Te\`orica and Institut de F\'{\i}sica Corpuscular,\\
Universitat de Val\`encia-CSIC, E-46100 Burjassot (Valencia), Spain.\\
vicente.vento@uv.es}

%\begin{history}
%\received{(received date)}
%\revised{(revised date)}
%%\accepted{(Day Month Year)}
%%\comby{(xxxxxxxxxx)}
%\end{history}

\begin{abstract}
Glueballs are particles whose valence degrees of freedom are gluons and therefore
in their description the gauge field plays a dominant role.
We review recent results in the physics of glueballs with the aim
set on phenomenology and discuss the possibility of finding them
 in conventional hadronic experiments and in the Quark Gluon
Plasma. In order to describe their properties we resort to a variety
of theoretical treatments which include, lattice QCD, constituent
models, AdS/QCD methods, and QCD sum rules. The review is supposed to be
an informed guide to the literature. Therefore, we do not discuss
in detail technical developments but refer the reader to the
appropriate references.
\end{abstract}
\maketitle
\section{Introduction}

Quantum Chromodynamics (QCD) is the theory of the hadronic interactions. It is an elegant
theory whose full non perturbative solution has escaped our knowledge since its formulation
more than 30 years ago.\cite{FritzschGellMannLeutwyler} The theory is asymptotically
free\cite{Gross:1973id,Politzer:1973fx} and confining.\cite{Wilson:1974sk} A particularly good
test of our understanding of the nonperturbative aspects of QCD is to study particles where
the gauge field plays a more important dynamical role than in the standard hadrons. In
particular glueballs, bound states of gluons, represent such a scenario.

The glueball spectrum has attracted much attention since the early days of
QCD.\cite{FritzschMinkowski} The interest in this subject is related to the significant
progress in the understanding of the properties of such states within QCD, as well as, in the
new possibilities for their identification in modern experiments. Historically the
investigation of the glueball properties started in the bag model by Jaffe and
Johnson.\cite{Jaffe:1975fd} They found many glueball states with different quantum numbers
lying in the mass interval 1000-2000 MeV. They emphasized that one should expect rather small
widths for such states because their decays in conventional hadrons violate the
Okubo-Zweig-Iizuka (OZI) rule.\cite{ozi}

After this pioneering work the study of glueballs was carried out by
using various versions of constituent models,
%\cite{Barnes:1981ac,Isgur:1984bm,Petersen,Brau:2004xw,Mathieu:2008bf,Kaidalov:1999yd}
by exploiting the QCD sum rule approach
%\cite{Novikov:1979va,Schafer:1994fd,Narison:1996fm,Forkel:2003mk}
and by performing lattice QCD calculations.
%\cite{Morningstar:1997ff,Morningstar:1999rf,Chen:2003vz,Bali,Weingarten,Lucini:2001ej,Hart}
Glueballs have not been an easy subject to study due to the lack of phenomenological support
and therefore much debate has been associated with their properties. The main achievement of
these approaches is the understanding of the deep relation between the properties of the
glueball states and the structure of the QCD vacuum. Besides, they provide a determination of
the spectrum both in gluodynamics, the theory with just gluons and no quarks, and in QCD.
However, in (unquenched) QCD, the results of several calculations for the spectrum are still
not universally accepted, in particular, for the lowest lying glueballs.\cite{Hart,Bali}

From the phenomenological point of view it has become clear by now that it is difficult to
single out which states of the hadronic spectrum are glueballs because we lack the necessary
knowledge to determine their decay properties. Moreover the strong expected mixing between
glueballs and quark states leads to a broadening of the possible glueball states which does
not simplify their isolation. The wishful sharp resonances which would confer the glueball
spectra the beauty and richness of the baryonic and mesonic spectra are lacking. This
confusing picture has led to a loss of theoretical and experimental interest in these hadronic
states. However, it is important to stress, that if they were to exist they would be a
beautiful and unique consequence of QCD. At the present, several candidates for the low mass
glueballs with quantum numbers $0^{++}$, $2^{++}$, $0^{-+}$ and $1^{--}$ are under discussion.
\cite{Klempt:2007cp,Ochs:2006rb,Anisovich:2005iv,Amsler:1995tu,Chanowitz:2006wf,Giacosa:2005bw}

In this review we will discuss the modern development in glueball spectroscopy from various
perspectives. In section~\ref{sec:lattice} we will summarize the results that lattice
techniques have obtained for the spectrum, both in the pure gauge theory and in the unquenched
calculations. In section~\ref{sec:models} we present a review on constituent models.
Section~\ref{sec:sumrules} is dedicated to discuss QCD sum rules. In
section~\ref{sec:prodecay} the production and decay mechanisms of glueballs in hadronic
reactions are discussed. In section~\ref{sec:qgp} the peculiarities of glueball production and
behavior in Quark-Gluon Plasma (QGP) is considered. Section~\ref{sec:other} is dedicated to
present two open topics the relation between the pomeron and glueballs and glueball-quarkonium
mixing. Finally in section \ref{sec:conclusions} we extract the main conclusions of our
analysis and try to foresee future developments.

\section{Lattice QCD}\label{sec:lattice}
\subsection{Overview}
Gluon self-couplings in QCD suggest the existence of glueballs,
bound states of mainly gluons. Investigating glueball physics
requires an intimate knowledge of the confining QCD vacuum and it is
well known that such properties cannot be obtained using standard
perturbative techniques. To handle the nonperturbative regime of
QCD, one can resort to numerical methods, known as lattice QCD.
Lattice QCD needs as input the quark masses and an overall scale,
conventionally given by $\Lambda_{QCD}$. Then any Green function can
be evaluated by taking average of suitable combinations of lattice
fields in vacuum samples. This allows masses  and matrix elements,
particularly those of weak or electromagnetic currents, to be
studied. One limitation of the lattice approach is in exploring
hadronic decays because the lattice, using Euclidean time, has no
concept of asymptotic states.

Lattice QCD was originally formulated by Wilson~\cite{Wilson:1974sk}
and is a clever implementation of the QCD dynamics using a finite
difference formalism. The starting point is the correlation function
in a discrete Euclidean space
\begin{equation}\label{eq:latt:action}
    C(t)=\langle\Omega|\phi^\dag(t)\phi(0)|\Omega\rangle\sim\int
     dU\int d\psi\int d\bar\psi \sum_{\bm x}
    \phi^\dag(\bm0,0)\phi(\bm x,t)e^{-S_F(\beta)-S_G(\beta)},
\end{equation}
where $S_F$ is the fermion action and $S_G$ the pure gauge action.
The continuum limit is controlled by the input parameter
$\beta=2N/g^2$ ($N$ is the number of color). By varying the inverse lattice coupling $\beta$ we
vary the lattice spacing $a$. The fermion action can be integrated
out exactly in Eq.~\eqref{eq:latt:action} to produce the fermion
determinant. The determinant describes the dynamics of the sea
quarks. In quenched QCD calculations, the determinant is set to a
constant.\cite{McNeile:2002en}

The physics is extracted from the fit
\begin{equation}\label{massexp}
    C(t) = \sum_{n}|\langle\Omega|\phi|n\rangle|^2\exp(-M_nt).
\end{equation}
$|n\rangle$ are the energy eigenstates and $M_n$ the corresponding masses. In practice, one
has to choose the operator $\phi$ which best overlaps with the lowest-lying glueball in the
channel of interest. $\phi$ is thus expanded in a basis with well-defined symmetry properties
under the octahedral group and the variational coefficients are determined by Monte-Carlo
simulations.\cite{Morningstar:1999rf}

The spectrum in a box with periodic boundary conditions includes not only single glueball
states, but also states consisting of several glueballs and torelons (gluon excitations which
wrap around the toroidal lattice). Fortunately, torelons are found to overlap only weakly with
single glueball states.\cite{Morningstar:1999rf,Meyer:2004jc}

Since  (classical) gluodynamics is dimensionless, its observables
will be also dimensionless. Masses are usually expressed in terms of
the string tension $m/\sqrt{\sigma}$ or the hadronic scale parameter
$r_0m$ defined through the static potential between quarks
$[r^2dV(r)/dr]_{r=r_0}=1.65$.\cite{Morningstar:1999rf,Chen:2005mg}
Their values are very close and usually read
\begin{align}\label{eq:latt:scale}
    \sqrt{\sigma} &= 440\pm20\text{ MeV,} &
    r_0^{-1} &= 410\pm20\text{ MeV.}
\end{align}
In lattice calculations there are errors arising from the finite
size of the lattice spacing $a$ and the finite lattice volume. But
for small enough $a$ we expect the continuum limit to be approached
as
\begin{equation}\label{eq:latt:err}
    \frac{m(a)}{\sqrt{\sigma(a)}} = \frac{m(0)}{\sqrt{\sigma(0)}}+ ca^2\sigma(0),
\end{equation}
with a constant $c$.\cite{Lucini:2001ej}

\subsection{Pure gauge spectrum}
The pure gauge spectrum of quenched QCD was
 investigated initially
by Morningstar and Peardon in an anisotropic
lattice.\cite{Morningstar:1999rf} They used different spacings for
the spatial $a_s$ and for the temporal $a_t$ extensions with the
ratio $\xi=a_s/a_t$. This technique allows to control better the
inherent errors induced by the lattice.

Morningstar and Peardon identified 13 glueballs below 4 GeV. In
order to distinguish single from multiple glueball states, they
determined approximately the locations of the two-particles states
using the mass estimation of the lowest few particles. In their
estimates of these locations they assumed that the two glueballs do
not interact and that the threshold for their production is given by
the energy
\begin{equation}\label{}
    E_{2G}\approx\sqrt{\bm p^2_1+m_1^2}+\sqrt{\bm p^2_2+m_2^2},
\end{equation}
where $\bm p_1=-\bm p_2$ and $m_{1,2}$ are the masses of the single
glueballs. All states lying below the two-glueball threshold are
then single glueball states. These authors pointed out that they
cannot rule out a single glueball interpretation for higher states.
They do not find any hint that their states are torelon pairs.

Finite volume effects are quite under control. When going from a
$6^3\times40$ lattice to a $8^3\times40$ lattice, the fractional
changes on the mass $\delta=1-m'/m$ are less than a few percent and
consistent with zero. The largest effect of these errors occur in
the $A_1^{*++}$ and $T_1^{*+-}$ representations of the octahedral
group. They are the main cause of  uncertainties in the $0^{*++}$
and $2^{*+-}$ glueballs. The proximity of the two glueball
thresholds and the finite volume errors on the $A^{*++}_1$ lead the
authors to withhold judgment on whether or not this level is a
single glueball.

\begin{figure}[t]
\begin{center}
\epsfig{file=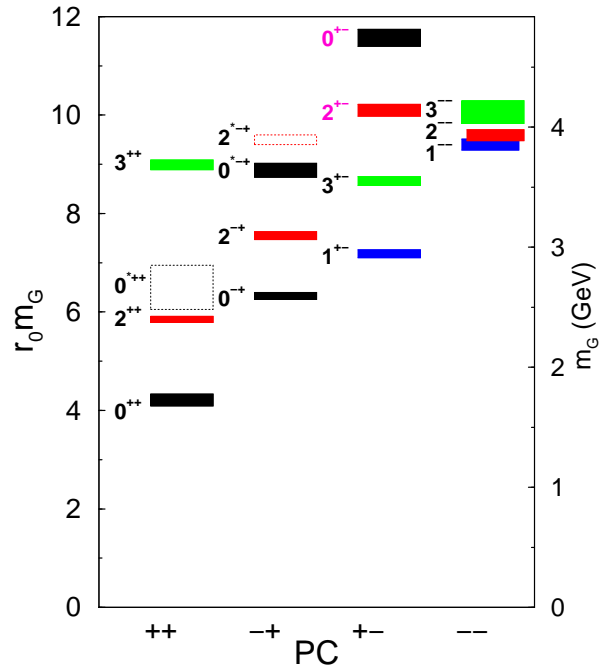,width=8cm}
\end{center}
\caption{The mass spectrum of glueballs in pure $\rm SU_C(3)$ gauge
theory . The masses are given in units of the hadronic scale $r_0$
along the left vertical axis and in GeV along the right vertical
axis. The mass uncertainties indicated by the vertical extents of
the boxes do not include the uncertainty in $r_0$. Numerical results
are listed in the table below. In some cases, the spin-parity
assignment for a state is not unique. The figure shows the smallest
$J$ value, the other possibilities are indicated in the second
column of the table.\label{fig:finalmasses}}
\end{figure}

\begin{table}[t]
\caption{
 Final continuum-limit glueball mass estimates $m_G$.
 When a unique $J$ interpretation for a state cannot
 be made, the other possibilities are indicated in the second column.
 States whose interpretation requires further study are indicated by a dagger.
 In column 3, the first error is the statistical uncertainty from the
 continuum-limit extrapolation and the second is the estimated uncertainty
 from the anisotropy. In the final column, the first error comes
 from the combined uncertainties in $r_0 m_G$, the second from the
 uncertainty in $r_0^{-1}=410(20)$ MeV.
\label{tab:finalmasses}}
\begin{center}
\begin{minipage}{3.0 true in}
\begin{tabular}{cccc}
\hline\hline
 $J^{PC}$ & Other $J$ & \hspace{1ex} $r_0 m_G$ & \hspace{2ex} $m_G$ (MeV) \\
\hline
 $0^{++}$  &                &  4.21 (11)(4)            & 1730 (50)(80)  \\
 $2^{++}$  &                &  5.85 (2)(6)          & 2400 (25)(120) \\
 $0^{-+}$  &                &  6.33 (7)(6)          & 2590 (40)(130) \\
 $0^{*++}$ &                &  6.50 (44)(7)$^\dagger$  & 2670 (180)(130)  \\
 $1^{+-}$  &                &  7.18 (4)(7)          & 2940 (30)(140) \\
 $2^{-+}$  &                &  7.55 (3)(8)          & 3100 (30)(150) \\
 $3^{+-}$  &                &  8.66 (4)(9)          & 3550 (40)(170) \\
 $0^{*-+}$ &                &  8.88 (11)(9)            & 3640 (60)(180) \\
 $3^{++}$  & $6,7,9,\dots$  &  8.99 (4)(9)          & 3690 (40)(180) \\
 $1^{--}$  & $3,5,7,\dots$  &  9.40 (6)(9)          & 3850 (50)(190) \\
 $2^{*-+}$ & $4,5,8,\dots$  &  9.50 (4)(9)$^\dagger$& 3890 (40)(190) \\
 $2^{--}$  & $3,5,7,\dots$  &  9.59 (4)(10)         & 3930 (40)(190) \\
 $3^{--}$  & $6,7,9,\dots$  &    10.06 (21)(10)           & 4130 (90)(200) \\
 $2^{+-}$  & $5,7,11,\dots$ &    10.10 (7)(10)         & 4140 (50)(200) \\
 $0^{+-}$  & $4,6,8,\dots$  &    11.57 (12)(12)           & 4740 (70)(230)\\
 \hline\hline
\end{tabular}
\end{minipage}
\end{center}
\end{table}

Lattice spacing errors, see Eq.~\eqref{eq:latt:err}, are expected to be ${\cal O}(a_t^2,a_s^4,
\alpha_sa_s^2)$ from perturbation theory.\cite{Morningstar:1999rf,Morningstar:1997ff} But the
results for different $\xi$ suggest that ${\cal O}(a^2_t,\alpha_sa_s^2)$ errors are
negligible. They extrapolate to the continuum limit assuming ${\cal O}(a_s^4)$. In addition to
these lattice errors of the dimensionless masses $r_0m$, one has to add the error arising from
the scale parameters~\eqref{eq:latt:scale} when presenting the absolute masses.

Representations of the octahedral group are distinct from
conventional spin representations of the Lorentz group. But one
expects that in the continuum limit, the former match onto the
latter. Once this extrapolation is achieved, one needs to identify
the quantum number $J^{PC}$ of the lattice spectrum. The low-lying
states do not lead to ambiguities. The situation is different for
higher states since they can belong to another multiplet with higher
excitation not calculated.

\begin{table}
\begin{center}
\caption{
  Glueball mass ratios. This ratios are not contaminated by
  anisotropy errors and are calculated using the empirical fact that
  correlations between different channels were found negligible.
  Note that the pseudoscallar glueball is clearly resolved to be
  heavier than the tensor.
\label{tab:massratios}} \vspace{0.2cm}
\begin{minipage}{2.0 true in}
\begin{tabular}{cc}
\hline\hline
$m(2^{++})/m(0^{++})$     &     $1.39(4)$ \\
$m(0^{-+})/m(0^{++})$     &     $1.50(4)$ \\
$m(0^{\ast++})/m(0^{++})$  &     $1.54(11)$ \\
$m(1^{+-})/m(0^{++})$     &     $1.70(5)$ \\
$m(2^{-+})/m(0^{++})$     &     $1.79(5)$ \\
$m(3^{+-})/m(0^{++})$     &     $2.06(6)$\\
$m(0^{\ast-+})$/$m(0^{++})$  &  $2.11(6)$ \\
$m(0^{-+})$ /$m(2^{++})$     &  $1.081(12)$ \\
\hline\hline
\end{tabular}
\end{minipage}
\end{center}
\end{table}

Their final results for the glueball spectrum are shown in
Fig.~\ref{fig:finalmasses} and in  Table~\ref{tab:finalmasses}. In
the figure they assume the most likely spin interpretations. The
table also contains any alternative spin attributions which cannot
be ruled out. Several mass ratios are shown in
Table~\ref{tab:massratios} which can be determined very accurately
since they are not contaminated by anisotropy errors. Note that the
pseudoscalar glueball is resolved to be heavier than the tensor.

To convert lattice glueball masses into physical units the value of
the hadronic scale $r_0$ must be specified. The estimate used,
$r_0^{-1}=410(20)$ MeV, was obtained by combining Wilson action
calculations of $a/r_0$ with values of the lattice spacing $a$
determined using quenched simulation results of various physical
quantities, such as masses of $\rho$ and $\phi$ mesons, the decay
constant $f_\pi$, and the $1P-1S$ splittings in charmonium and
bottomonium. A great deal of care should be taken in making direct
comparisons with experiments since these values neglect the effects
of light quarks and mixings with nearby conventional mesons.

More recently Chen et al.\cite{Chen:2005mg} have performed a similar calculation with larger
lattices and larger volumes. We present in the Table~\ref{tab:comparison} a comparison of
their results with those of Morningstar and Peardon.\cite{Morningstar:1999rf}

\begin{table}
\begin{center}
\caption{
  Continuum-limit glueball masses $M_G$ for Chen et al.
  and for Morningstar and Peardon. \label{tab:comparison} } \vspace{0.2cm}
\begin{tabular}{cccc}
$R^{PC}$  &  Possible $J^{PC}$& $r_0 M_G$~\cite{Chen:2005mg} & $r_0
M_G$~\cite{Morningstar:1999rf}\\
\hline\hline
$A_1^{++}$&  $0^{++}$ & 4.16(11)   &    4.21(11)    \\

$E^{++}  $&  $2^{++}$ & 5.82(5)    &    5.85(2)     \\
$T_2^{++}$&  $2^{++}$ & 5.83(4)    &    5.85(2)     \\

$A_2^{++}$&  $3^{++}$ & 9.00(8)    &    8.99(4)     \\
$T_1^{++}$&  $3^{++}$ & 8.87(8)    &    8.99(4)     \\

$A_1^{-+}$&  $0^{-+}$ & 6.25(6)    &    6.33(7)     \\

$T_1^{+-}$&  $1^{+-}$ & 7.27(4)   &    7.18(3)     \\

$E^{-+}  $&  $2^{-+}$ & 7.49(7)    &    7.55(3)     \\
$T_2^{-+}$&  $2^{-+}$ & 7.34(11)   &    7.55(3)     \\

$T_2^{+-}$&  $3^{+-}$ & 8.80(3)   &     8.66(4)     \\
$A_2^{+-}$&  $3^{+-}$ & 8.78(5)   &    8.66(3)     \\

$T_1^{--}$&  $1^{--}$ & 9.34(4)   &    9.50(4)     \\

$E^{--}$&    $2^{--}$ & 9.71(3)   &    9.59(4)     \\
$T_2^{--}$&  $2^{--}$ & 9.83(8)   &    9.59(4)     \\

$A_2^{--}$&  $3^{--}$ & 10.25(4)   &    10.06(21)     \\

$E^{+-}$&    $2^{+-}$ & 10.32(7)   &    10.10(7)     \\

$A_1^{+-}$&  $0^{+-}$ & 11.66(7)   &    11.57(12)     \\
\hline\hline
\end{tabular}
\end{center}
\end{table}

\begin{figure}[htb]
\begin{center}
\epsfig{file=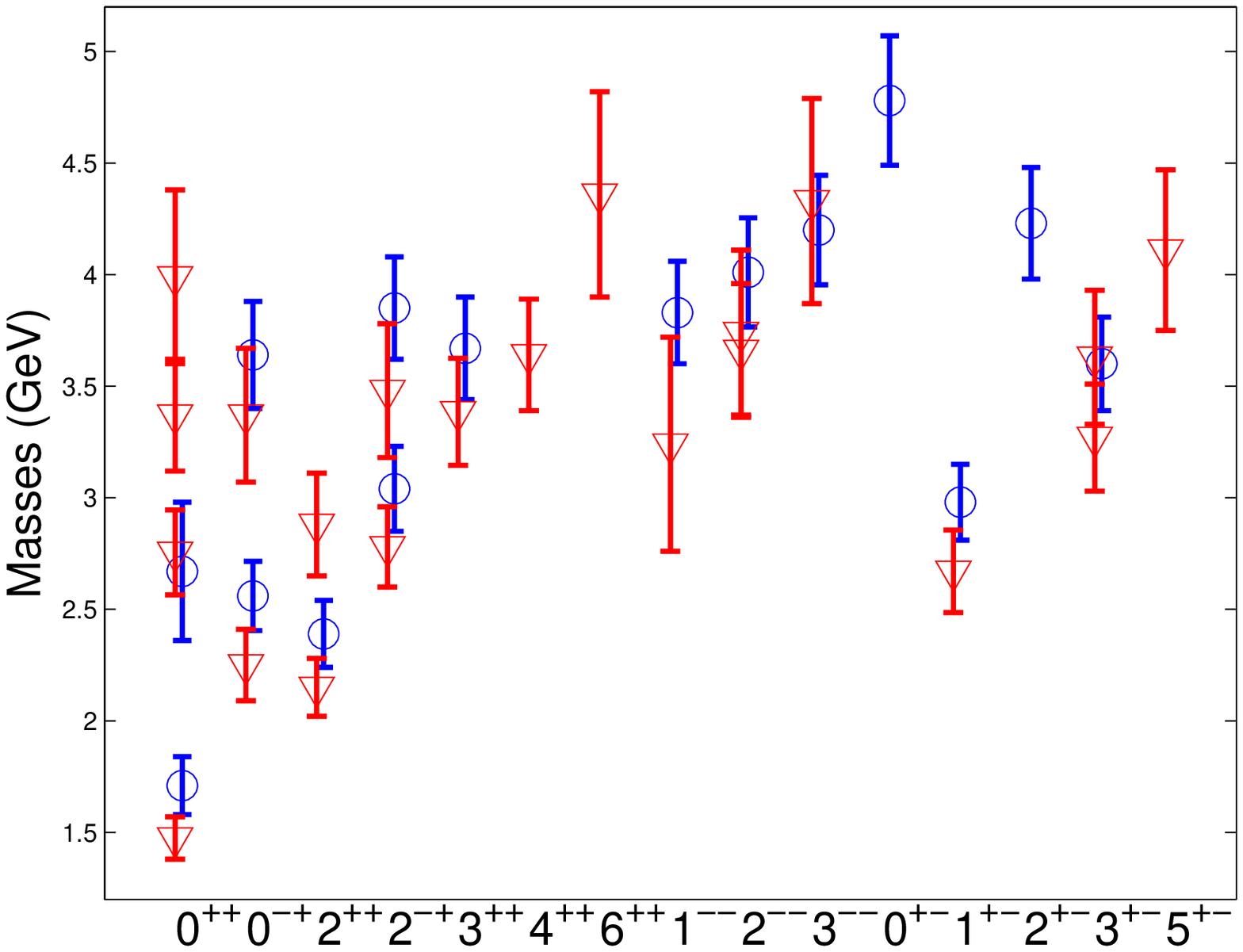,width=0.46\linewidth}\hspace{0.5cm}
\epsfig{file=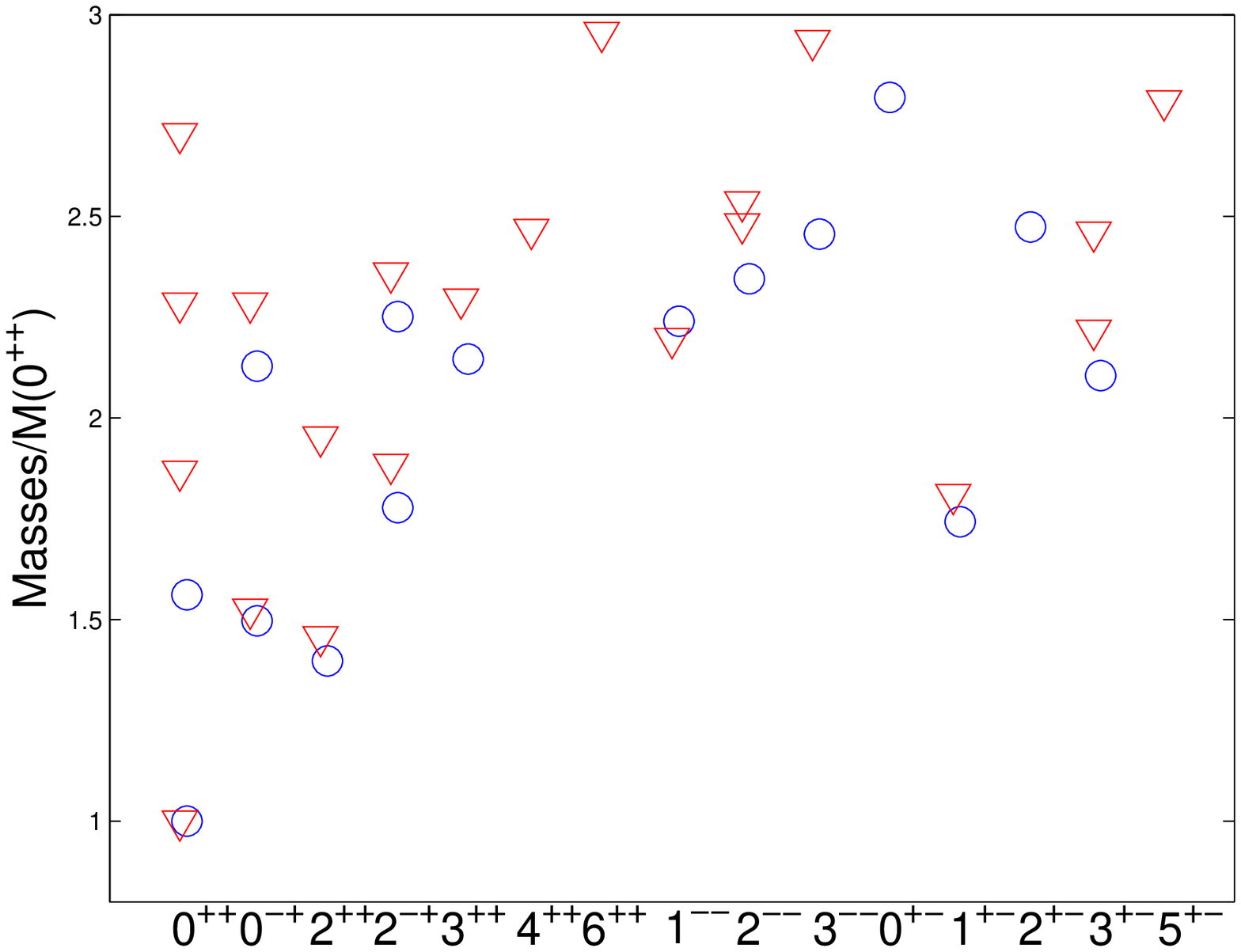,width=0.47\linewidth}
\end{center}
\caption{Comparison between Morningstar and Peardon (circles) and Meyer and Teper (triangles)
of the mass spectrum of glueballs in pure $\rm SU_C(3)$ gauge theory. Absolute masses (left)
and mass ratios with respect the scalar glueballs (right). \label{fig:comp_latt}}
\end{figure}

Meyer and Teper investigated also the pure gauge spectrum for (even)$^{++}$ states on a
lattice in order to check the linearity of the Pomeron trajectory (see
section~\ref{sec:other}).\cite{Meyer:2004jc} They reported also masses in other $PC$ sectors.
It is instructive to compare their results with the  Morningstar and Peardon
study.\cite{Morningstar:1999rf} Although in these works absolute masses are expressed with
different energy scales ($r_0^{-1}$,\cite{Morningstar:1999rf} and
$\sqrt{\sigma}$,\cite{Meyer:2004jc}), their close values [see Eq. \eqref{eq:latt:scale}]
 allow the comparison of their absolute spectra displayed in
Fig.~\ref{fig:comp_latt} (left). Level orderings in both cases are
identical but globally Meyer and Teper masses are smaller.
Fig.~\ref{fig:comp_latt} (right) presents mass ratios which are not
contaminated by anisotropy errors. In this case, errors bars are of
the order of the symbols and are not shown.

Recently, Meyer updated masses of the scalar and the tensor using his
technique.\cite{Meyer:2008tr} In this latter reference, he used the lattice scale $r_0$
allowing to compare with the other studies. The new masses are closer the Morningstar and
Peardon's ones and read
\begin{align}\label{eq:MeyerMass}
    r_0M_{0^{++}}&=3.958(47), & r_0M_{2^{++}}&=5.878(77).
\end{align}
All lattice calculations are now consistent and shown that, in pure gauge theory, the masses
of the lowest states are
\begin{align}\label{eq:MeyerMass}
    M_{0^{++}}&\sim1.6-1.7 \text{ GeV}, & M_{2^{++}}&\sim2.4\text{ GeV
    }.
\end{align}

In full QCD interpolating operators  for a  state
with given quantum numbers can also be constructed from quarks
and anti-quarks. The pure glue operators might mix with the
fermionic operators. If the mixing is very strong, the glueball masses obtained
in this way, will have little to do with the glueball masses in the quenched calculation. Several
methods have been applied to the interesting scalar sector,
$J^{PC}=0^{++}$, of the physical spectrum.

Weingarten and Lee \cite{Lee:1999kv}   studied the effect of the effect of quarkonium mixing
with the glueball in the lowest $0^{++}$ state in quenched QCD. The results were expressed
as a mixing matrix
 \begin{equation} \left(
\begin{array}{c c}
m_g & E(s) \\
E(s) & m_{\sigma} (s) \\
\end{array} \right)
\end{equation}
where $m_g$ is the glueball mass, $m_\sigma(s)$ is the mass of the $0^{++}$ non-singlet  $\bar{\psi} \psi$ state,
and $E(s)$ is the mixing energy. Weingarten and Lee measured: $m_g$ = 1648(58) MeV,
$m_\sigma(s)$ = 1322(42) MeV, and $E(s)$ = 61(58) MeV. The qualitative picture that emerges
is that the $f_0$ (1710) is “mostly” $0^{++}$ glueball, and the $f_0(1500)$ is “mostly” $\bar{s}s$.
The mixing energy $E(s)$ has large lattice spacing errors. For example at a lattice spacing
of $a^{-1} \sim 1.2$ GeV, the Weingarten and Lee  result is $E(s)\sim$ 0.36 GeV,
while that of McNeile and Michael\cite{McNeile:2000xx}  is $E(s)\sim$ 0.44 GeV.
The analysis of Weingarten and Lee depends on the $0^{++}$ states being well defined
in quenched QCD. Bardeen et al.\cite{Bardeen:2001jm} have shown that there is a problem
 with the nonsinglet $0^{++}$
correlator in quenched QCD. The problem can be understood using quenched chiral perturbation
theory. The non-singlet $0^{++}$ propagator contains an intermediate state of $\eta^\prime -
\pi$ . The removal of fermion loops in quenched QCD has a big effect on the $\eta^\prime$
propagator. The result is that a ghost state contributes to the scalar correlator, that makes
the expression in Eq. \eqref{massexp} inappropriate to extract masses from the calculation.

A lattice QCD calculation that includes the dynamics of the sea
quarks should reproduce the physical spectrum. Some insight into the
composition of individual physical states, such as wether the
physical particles couple to fermionic operators or pure operators
could be studied as an effect of decreasing sea quark mass. Some
studies have been performed for $n_f = 2$ QCD \cite{Hart} and it was
found that the mass of the $0^{++}$ glueball was reduced with
respect to the quenched calculation by about $20\%$. Not so the
tensor $2^{++}$, whose value remained close to the quenched
calculation. The mass of the $0^{++}$ glueball on the UKQCD data set
is degenerate with the mass of two pions.\cite{Hart} Due to the
intricacies of the physical spectrum the lattice spacing used in the
unquenched calculations must be reduced before direct contact can be
made to phenomenology.

In the real world glueballs have a decay width since they decay into
two mesons. Lattice QCD calculations are performed in Euclidean
space and this makes the computation of intrinsically complex
quantities, as decay widths, complicated.\cite{Michael:1989mf} The
decay width for the $0^{++}$ glueball to two pseudoscalars has been
calculated to be $108(28)$ MeV.\cite{Sexton:1995kd} It is
encouraging that it is small compared with its mass, however there
is not yet consensus in this result.\cite{Burakovsky:1998zg}

\section{Constituent Models}\label{sec:models}

Gluon self-coupling in QCD suggests the existence of glueballs. Incontrovertible experimental
evidence for their existence remains elusive. A primary reason for this is the difficulty in
extracting the properties of glueballs from the QCD lagrangian. We have seen that lattice QCD
faces both computational and fundamental problems. We next describe a complementary way to
describe glueballs, namely constituent models, which implement in a dynamical way the
phenomenological properties of the confining QCD vacuum and the interaction among the gluons.

\subsection{The MIT bag model}

Hadrons are physical systems where quarks and gluons are confined in
regions  smaller than 1~fm. This experimental fact led the
physicists of the MIT to develop a bag model of hadrons in
QCD.\cite{Chodos:1974je} In their picture, quarks and gluons are
confined in a bag, usually taken as a static spherical cavity.
Confinement is described in the model by a boundary condition and a
constant energy density $B$. The boundary condition makes the color
flux vanish at the surface of the bag and it produces a quantization
of the energy levels. $B$ gives a constant energy term which
stabilizes the bag at a finite size. Energy modes $E_i=x_i/R$ are
inversely proportional to the radius of the spherical cavity $R$.
The energy in the cavity under these conditions, with $n_i$ massless
constituents of type $i$, is
\begin{equation}\label{eq:bag1}
    E =  \frac{4\pi BR^3}{3} + \sum_i n_i\frac{x_i}{R}.
\end{equation}
The bag energy~\eqref{eq:bag1} encodes the masses of the states $M$
but also the dynamics of the center-of-mass motion. One solution to
this problem is the following: We consider that the bag is an
eigenstate of the Hamiltonian $H^2=\bm P^2+M^2$. Then we simply take
out the quantity
\begin{equation}\label{}
    \left\langle P^2\right\rangle = \sum_i n_i\left(\frac{x_i}{R}\right)^2.
\end{equation}
The mass then reads
\begin{equation}\label{}
    M^2 = E^2- \left\langle P^2\right\rangle.
\end{equation}
In their application to  hadrons,\cite{Chodos:1974pn} the authors
considered the bag constant as a free parameters.  The minimization
of the mass equation leads to a relation between the radius and $B$.
Then the mass equation can be written in terms of only $B$ and by
fitting hadronic masses one can determine this parameter.

Jaffe and Johnson were the first to apply this model to
glueballs.\cite{Jaffe:1975fd} They found the modes of the gluon
field in the cavity corresponding to the solution of the gluon
equations of motion subject to the boundary conditions
\begin{equation}\label{eq:bag_bound_cond}
    n_\mu G^{\mu\nu}_a = 0,
\end{equation}
where $n_\mu$ is the normal to the bag surface
and $G^{\mu\nu}_a $ is the gluon field strength. The two lowest modes
are:
 \begin{align}
    \text{Transverse Electric}&& J^P=&1^+ & x_{\text{TE}}&=2.744,\\
    \text{Transverse Magnetic}&& J^P=&1^-& x_{\text{TM}}&=4.493.
\end{align}
From these one obtains immediatly the masses of the low-lying
states: (TE)$^2$, $0^{++}, 2^{++}$, $M=960$ MeV ; (TE)(TM), $0^{-+},
2^{-+}$, $M=1.3$ GeV ; (TE)$^3$, $0^{++}, 1^{+-}, 3^{+-}$, $M=1.45$
GeV. Some authors pointed out the fact that lowest three-gluon
glueballs $0^{++}, 1^{+-}, 3^{+-}$ have the opposite parity to that
of the potential model predictions.\cite{Donoghue:1980hw} The parity
of the TE mode $1^+$ causes this difference.  The lattice spectrum
seems to support the bag model.

Differently to potential models, constituents in the bag model are not confined by a
potential. Particles are almost free inside the bag and the confinement is ensured by the
boundary conditions~\eqref{eq:bag_bound_cond}. A more elaborated description should lift
degeneracies. Carlson {\it et al.} \cite{Carlson:1982er} added to the bag
energy~\eqref{eq:bag1} a correction $\Delta E$ representing the spin splitting induced by
one-gluon-exchange interaction. This shift includes tree-level scattering diagrams but also
self-energy contributions,
\begin{subequations}\label{eq:bagsplit}
\begin{eqnarray}
  \Delta E &=& \sum_{i\neq j} \Delta E_{ij} + \sum_{i} \Delta E_{i},\\
  \Delta E_{ij} &=& -\frac{\alpha(R)}{R}\langle\bm t_i\cdot\bm t_j (a_{ij}\bm S_i\cdot\bm S_j
  + b_{ij} T_{ij} + c_{ij} I_{ij}) + d_{ij} {\cal P}_{ij}\rangle, \\
  \Delta E_{i} &=& -\frac{\alpha(R)}{R} \bm t_i^2 e_i,
\end{eqnarray}
\end{subequations}
Here $\bm t_i$ and $\bm S_i$ are the generators of color and spin
respectively. ${\cal P}$ is the projector onto the color-octet
spin-one state. $I$ is the identity operator and $\alpha(R)$ is the
running coupling constant,

\begin{equation}\label{}
    \alpha(R) = \frac{2\pi n}{9}\frac{1}{\ln[1+1/(\Lambda R)^n]}.
\end{equation}
This Ansatz for $\alpha(R)$ simulates the saturation for large $R$.
$n$ is a positive integer parameter. The authors used $n=2$ and
$\Lambda=0.172$ GeV in their calculations. The coefficients $a, b,
c, d, e $ in the relations~\eqref{eq:bagsplit} are given in the
reference.

\begin{figure}[htb]
\begin{center}
\epsfig{file=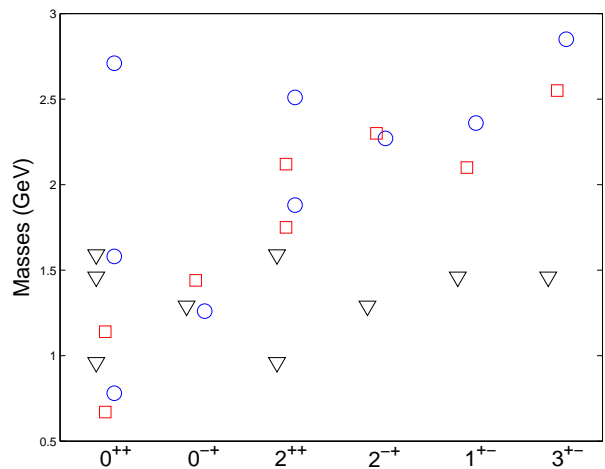,width=8cm}

\end{center}
\caption{The mass spectrum of glueballs in bag models: Jaffe and Johnson$^6$ (triangles),
Carlson {\it et al.}$^{32}$ (blue circles) and Chanowitz and Sharpe$^{35}$
(squares).\label{fig:bagmasses}}
\end{figure}

In the original version of the bag model, the bag constant was a
free parameter determined from data. Hansson {\it et al.} proposed a
model for the QCD vacuum wich allows B to be calculated given
$\alpha(R)$ and $e_{\text{TE}}$ (the self-energy of the lowest TE
gluon mode).\cite{Hansson:1982dv} The basic idea is that the vacuum
is filled with $0^{++}$ (TE)$^2$ glueballs which form a negative
energy condensate. The expression for $B$
\begin{equation}\label{}
    B = \frac{3}{8\pi R_o^3}(-m^2)^{1/2}
\end{equation}
relates its value to the (negative) mass squared of the scalar glueballs $m^2$. The excitation
of this condensate gives rise to a observable scalar glueballs.\cite{Carlson:1984wq} Their
results are shown in Fig.~\ref{fig:bagmasses}. This figure displays also the glueball spectrum
obtained in the bag model of Chanowitz and Sharpe along similar lines.\cite{Chanowitz:1982qj}
Masses in the bag model are globally lower than in lattice QCD.

In bag models, particles are confined in a spherical bag. This
approximation is only valid for totally symmetric $J=0$ states.
Moreover, spherical glueballs are not stable since a vector field
can never give rise to a spherically symmetric pressure. Robson
pointed out this flaw and developed a toroidal bag model for
glueballs.\cite{Robson:1978iu} This picture of glueballs is close to
the flux tube model of Isgur and Paton.\cite{Isgur:1984bm}

\subsection{Gluon mass}

A possible way to handle glueballs is to consider massive
quasi-gluons interacting via a QCD inspired dynamics. The gauge
bosons are massless at the Lagrangian level but there are hints that
they obey massive dispersion relations. The gluons remains massless
to all orders in perturbation theory. However, nonperturbative
effects, like confinement, and their self-interactions, can be
described by a constituent mass. Several definitions exist for this
constituent mass.

The so called dynamical mass, is defined by the position of the pole
of the dressed gluon propagator. Cornwall arrived to such a
dynamical mass by analyzing the gluon Dyson-Schwinger equations in
the early 80's.\cite{Cornwall:1981zr} This infinite set of couple
equations cannot be solved analytically. One must resort to a
truncation scheme. By a clever resummation of Feynman diagrams,
Cornwall found a gauge-invariant procedure to deal with these
equations. With this technique, a full gluon propagator in quarkless
QCD emerges
\begin{equation}\label{eq:gluon_prop_cornwall}
d^{-1}(q^2) = \left(q^2+m^2(q^2)\right) bg^2 \ln\left[\frac{q^2+4m^2(q^2)}{\Lambda^2}\right],
\end{equation}
with a dynamical mass
\begin{equation}\label{eq:dym_mass}
    m^2(q^2) =
    m^2\left(\frac{\ln\left[(q^2+4m^2)/\Lambda^2\right]}
    {\ln\left(4m^2/\Lambda^2\right)}\right)^{-12/11}.
\end{equation}
In Eq.~\eqref{eq:gluon_prop_cornwall},  $b=11N/48\pi^2$ is the first
coefficient of the beta function for quarkless QCD and the mass term
that appears has finite value at zero momentum. The gluon mass can
be related to the gluon condensate $\langle G_{\mu\nu}^a
G^{\mu\nu}_a\rangle$ from which the value $m=(500\pm200)$~MeV
arises.

Bernard proposed a different definition for the gluon
mass.\cite{Bernard:1981pg} Consider the potential energy of a pair
of heavy, static sources in the adjoint representation of the color
group. As the separation of the adjoint sources (static gluons) is
increased, the potential will increase linearly as a string or a
flux tube is  formed between them. The energy stored in the string
will at some point be large enough to pop up a pair of dynamical
gluons out of the vacuum. The effective gluon mass is defined as
half of the energy stored in the flux tube at this point.
Monte-Carlo simulations of this phenomenon show a effective gluon
mass in the range 500-800~MeV.

The effective gluon mass was also investigated in the bag model.\cite{Donoghue:1983fy} Even
though the gluon is massless in the bag model, a net energy of $740 \pm 100$~MeV is required
to produce a gluon due to confinement.

All these arguments support the use of an effective gluon mass to
describe the dynamics of QCD. It is therefore possible to envisage
an approach to bound states made of {\it constituent} massive
gluons. Since two-gluon glueballs have always a positive conjugation
charge, a study of the full spectrum must include also three-gluon
glueballs.

\subsection{Two-gluon glueballs}
One of the pioneering works on two-gluon glueballs was the study by Cornwall and
Soni.\cite{Cornwall:1982zn} The large value of the effective gluon mass led them to propose a
nonrelativistic approach to gluonium. They used a confining potential which saturates at
large distances constrained by Bernard's results\cite{Bernard:1981pg}
\begin{equation}\label{eq:pot_sat}
    V_{C}(r) = 2m\left(1-e^{-r/r_s}\right).
\end{equation}
This screened potential led to a spectrum with relatively low
glueball masses. The scalar and tensor glueballs had masses around
$\sim 1.3$ and $\sim 1.6$~GeV, respectively. They used for the
string tension twice the value commonly used for mesons, a decision
which they justified by arguing that a gluon acts as a $q\bar q$
pair. The Coulomb and spin-dependent interactions at short-range
were derived from a nonrelativistic expansion of the Feynman graphs
for two-gluon scattering. They considered massive exchanged gluons
(with the same mass as the constituent one) to keep the gauge
invariance of the amplitudes to the given order. This one-gluon
exchange (OGE) potential involves Yukawa, spin-orbit, spin-spin and
tensor forces,

\begin{equation}\label{eq:pot_oge_cornwall}\begin{split}
V_{\text{oge}}(\bm r) =& -\frac{\lambda e^{-mr}}{r}\left(\frac{2s-7m^2}{6m^2}+\frac{1}{3}\bm
S^2\right) + \frac{\pi\lambda\delta(\bm r)}{3m^2} \left(\frac{4m^2-2s}{m^2}+\frac{5}{2}\bm
S^2\right) \\
&- \frac{3\lambda}{2m^2}\bm L\cdot\bm
S\frac{1}{r}\frac{\partial}{\partial r} \frac{e^{-mr}}{r} +
\frac{\lambda}{2m^2}\left[\left(\bm S\cdot\bm\nabla\right)^2
-\frac{1}{3}\bm S^2\bm\nabla^2\right]\frac{e^{-mr}}{r}.
\end{split}
\end{equation}
$s$ is the glueball mass squared, which we can set to $s=4m^2$ in a
first approximation, and $\lambda=3\alpha_s$ is the strong coupling
constant.

Cornwall and Soni  presented results for states with quantum numbers
$L=0$, $J^{PC} = 0^{++}, 2^{++}$, and $L=1$, $J^{PC} = 1^{-+},
2^{-+}$. Despite the fact that the gluons acquire a mass they remain
transverse. For transverse particles the $J=1$ states are forbidden
as a consequence of Yang's theorem,\cite{Yang:1950rg} {\it i.e.}, a
gluon in a massless representation has only two projections for its
spin and therefore two transverse gluons cannot bind into a $J=1$
state. Thus we must incorporate this feature into the above
formalism.

When dealing with massless representations, the conventional $\bm J=\bm L+\bm S$ decomposition
is not useful anymore. The formalism to treat two-body relativistic scattering developed by
Jacob and Wick\cite{Jacob:1959at} allows also the description of representations with only
transverse gluons. We sketch its main features and apply this formalism to the study of the
two-gluon glueball.

The Jacob and Wick formalism is based on states, $\left|J,M;\lambda_1,\lambda_2\right\rangle$,
which are eigenstates of $\bm J^2$ and $J_z$ and where $\lambda_1$ and $\lambda_2$ represent
the two allowed spin projections. In the case under consideration, the projections can only be
maximal, {\it i.e.} $\pm s$ for a particle with helicity-$s$ . The angular part of these
states are related with the conventional basis states by means of Clebsch-Gordan coefficients:
\begin{equation}\label{eq:decomp}
    \left|J,M;\lambda_1,\lambda_2\right\rangle=\sum_{L,S}
    \left[\frac{2L+1}{2J+1}\right]^{1/2}
    \left\langle LS0\Lambda\right|J\Lambda\left.
    \right\rangle\left\langle s_1s_2\lambda_1(-\lambda_2)
    \right|S\Lambda\left.
    \right\rangle\, \left|^{2S+1}L_J\right\rangle,
\end{equation}
with $\Lambda=\lambda_1-\lambda_2$. The radial part (depending on $J$) is determined
variationally with the Hamiltonian. The states are not eigenstates of parity. For a two-gluon
state, $s_1=s_2=1$, it holds

\begin{equation}\label{pdef}
    \hat P\left|J,M;\lambda_1,\lambda_2\right\rangle=(-1)^{J}
    \left|J,M;-\lambda_1,-\lambda_2\right\rangle.
\end{equation}
The key point is that physical states must not only be eigenstates of the total angular
momentum operator but they must also be eigenstates of parity. Such a requirement is fulfilled
by the following linear combinations
\begin{equation}\label{hdef0}
    \left|H_\pm; J^P;\lambda_1,\lambda_2\right\rangle=\frac{1}{\sqrt 2}\left\{\, \left|J,M;\lambda_1,\lambda_2\right\rangle
    \pm\left|J,M;-\lambda_1,-\lambda_2\right\rangle\right\},
\end{equation}
for which $\textsf{P}\left|H_\pm; J^P;\lambda_1,\lambda_2\right\rangle=P \left|H_\pm;
J^P;\lambda_1,\lambda_2\right\rangle$, with $P=\pm(-1)^{J}$. In the latter, the $\left|H_\pm;
J^P;\lambda_1,\lambda_2\right\rangle$ states will be referred
as helicity states. When the two particles are identical, the wave function should be an
eigenvector of the permutation operator ${\mathrm P}_{12}$. The basic state~\eqref{hdef0} are
eigenstate of the permutation in the case of massless particles since $\lambda_1=\pm\lambda_2$
and\cite{Jacob:1959at}
\begin{equation}\label{symdef}
{\mathrm P}_{12}\left|J,M;\lambda_1,\lambda_2\right\rangle=
(-1)^{J}\left|J,M;\lambda_2,\lambda_1\right\rangle.
\end{equation}
A system of two gluons has to be totaly symmetric and this constraint leads to selection rules
on the spin and the parity. Indeed, for singlet states (defined by
$\lambda_1=\lambda_2=\lambda$), we have
\begin{equation}\label{}
{\mathrm P}_{12}\left|H_\pm; J^P;\lambda,\lambda\right\rangle= (-1)^{J} \left|H_\pm;
J^P;\lambda,\lambda\right\rangle,
\end{equation}
leading to the families with the even spin and positive or negative parity . The minimum spin
in this case is $J=0$. For the doublet (defined by $\lambda_1=-\lambda_2=\lambda$), the
minimum spin is $J=2$ and we have
\begin{equation}\label{}
{\mathrm P}_{12}\left|H_\pm; J^P;\lambda,-\lambda\right\rangle= \pm(-1)^{J} \left|H_\pm;
J^P;\lambda,-\lambda\right\rangle.
\end{equation}
This time we can have odd spin (the lowest is $3^+$) but also even spin with positive parity.
We thus observe the emergence of selection rules according to the value of $J$ and $P$.

The results for states of two transverse gluons are formally identical to the case of states
made of two photons, {\it i.e.}, Yang's theorem.\cite{Yang:1950rg} The total color wave
function is assumed to be a singlet, which is totally symmetric, and does not explicitly
appear in the computations. Taking into account the Bose symmetry, one finds that there are
four allowed helicity states, namely
\begin{subequations}\label{hsdef}
\begin{align}
    \left|S_\pm;J^P\right\rangle&=\left| H_{\pm};J^P\right\rangle_{\lambda_2=\lambda_1}, \\
    \left|D_\pm;J^P\right\rangle&=\left| H_{\pm}; J^P\right\rangle_{\lambda_2=-\lambda_1}.
    %\left|D_-;J^P\right\rangle&=\left| H_{-,-1}; J^P\right\rangle_{\lambda_2=-\lambda_1}.
\end{align}
\end{subequations}
The selection rules impose restrictions to the possible values of the total angular momentum
and parity of these four types of states. It can be checked that one can only obtain the
following states
\begin{equation}\label{ggstate}
        \left|S_+;(2k)^+\right\rangle,\quad \left|S_-;(2k)^-\right\rangle,\quad
        \left|D_+; (2k+2)^+\right\rangle,\quad \left|D_-; (2k+3)^+\right\rangle,
        \quad k\in\mathbb{N}.
\end{equation}
The $S$- and $D$-labels stand for helicity-singlet and -doublet
respectively. We recognized in Eq.~\eqref{ggstate} the four families
predicted by Yang.\cite{Yang:1950rg}

It should be noted that a state made of two gluons in a color
singlet state has always positive charge conjugation ($C=+1$). More
explicitly, the states in Eq.~(\ref{ggstate})  give rise to the
following glueball states
\begin{subequations}\label{ggstate3}
\begin{eqnarray}
    \left|S_+\; ; \; (2k)^+ _{}\;\right\rangle \;\;&\Rightarrow& 0^{++}, 2^{++}, 4^{++},\dots\\
    \left|S_-\; ; \; (2k)^-\;\right\rangle \;\;&\Rightarrow&0^{-+}, 2^{-+}, 4^{-+},\dots\\
    \left|D_+;(2k+2)^+\right\rangle&\Rightarrow& 2^{++},4^{++},\dots\\
        \left|D_-;(2k+3)^+\right\rangle&\Rightarrow&3^{++},5^{++},\dots
\end{eqnarray}
\end{subequations}
It is readily observed that only the
$\left|S_\pm;(2k)^+\right\rangle$ states can lead to $J=0$, while
the $\left|D_\pm\right\rangle$ states always have $J\geq 2$ (since
$J>|\lambda_1-\lambda_2|$). Obviously, a consequence of Yang's
theorem is that no $J=1$ states are present. Only the
$\left|D_-\right\rangle$ states can generate an odd-$J$, but $J$ is
at least 3 in this case.

Lattice QCD confirms the absence of the $1^{-+}$ and $1^{++}$
states, at least below $4$ GeV. It is worth mentioning that glueball
states with even-$J$ and positive parity can be built either from
the helicity-singlet or from the helicity-doublet states. The
important result is that the gluons remain transverse and therefore
the helicity formalism exactly reproduces the $J^{PC}$ content for
glueballs which is observed in lattice QCD, without the extra states
which are usually present in potential models.

The helicity formalism was applied for the first time by
Barnes.\cite{Barnes:1981ac} It has several advantages not shared by
the more conventional nonrelativistic $LS$-basis. It avoids
spurious states forbidden by the coupling of two transverse gluons
but also reproduces the lattice QCD hierarchy, {\it i.e.}
\begin{equation}\label{eq:latt_hierarchy}
0^{\pm+},2^{++},2^{\pm+},3^{++},4^{++},4^{\pm+},5^{++},6^{++}.
\end{equation}
Within this approach, a given $J^{PC}$ state can be expressed as a linear combination of
$(L,S)$ states thanks to Eq.~\eqref{eq:decomp}. The complete expressions for these
decompositions can be found in Mathieu {\it et al}.\cite{Mathieu:2008bf} We give here the
angular dependence of the ground states of Eq.~\eqref{ggstate3}:
\begin{subequations}
\begin{eqnarray}
\left|S_+;(0)^+\right\rangle&=&\sqrt{\frac{2}{3}}\left|^1
S_{0}\right\rangle +\sqrt{\frac{1}{3}}\left|^5 D_{0}\right\rangle,\\
    \left|S_-;(0)^-\right\rangle&=&\left|^3 P_{0}\right\rangle,\\
\left|D_+;(2)^+\right\rangle&=&\sqrt{\frac{2}{5}}\left|^5S_{2}\right\rangle
+\sqrt{\frac{4}{7}}\left|^5 D_{2}\right\rangle+\sqrt{\frac{1}{7}}\left|^5 G_{2}\right\rangle,\\
        \left|D_-;(3)^+\right\rangle&=&\sqrt{\frac{5}{7}}\left|^5D_{3}\right\rangle
    +\sqrt{\frac{2}{7}}\left|^5 G_{3}\right\rangle.
\end{eqnarray}
\end{subequations}
These decompositions are essential for computing the matrix elements
of relativistic operators (spin-spin, spin-orbit and tensor). Let us
note that the matrix elements of these operators are equal for
$\left|S_+;(2k)^+\right\rangle$ and
$\left|S_-;(2k)^-\right\rangle$.\cite{Mathieu:2008bf}

Even though in this approach the singlet states
$J^{P}=(2k^{+},2k^{-})$ are degenerate, with a  Cornell-type (linear
+ Coulomb) potential, a nonrelativistic kinetic energy, which
incorporates an {\it ad hoc} gluon mass $m$, and using the helicity
formalism, Barnes was able to reproduce the qualitative feature of
the pure gauge sector finding $M(0^{\pm+})=4.36 \, m$. The higher mass
ratios were not in perfect agreement with modern lattice results,
implying the need for modifications.

We emphasize that considering transverse gluons is essential for
finding the correct hierarchy. However, transverse particles are
generally massless and even if nonperturbative effects are able to
give a mass to the gluon, one may wonder if a nonrelativistic
kinetic energy $\bm p^2/2m$ is appropriate. Indeed, the
nonrelativistic kinetic energy is just the limit of the more
general Dirac operator $\sqrt{\bm p^2+m^2}$. This semi-relativistic
energy is also valid for massless particles such as gluons.

Brau and Semay compared different models for glueballs.\cite{Brau:2004xw} Models with a
nonrelativistic kinetic energy were not able to reproduce correctly the lattice gauge
spectrum for realistic values of the parameters. They concluded that a semi-relativistic
Hamiltonian, {\it i.e.} $2\sqrt{\bm p^2}+V(r)$, is an essential ingredient  to handle glue
states. Nevertheless, all models analyzed used an $LS$-basis and were plagued with unwanted
states. Moreover, they had to include the short-range potential Eq.~\eqref{eq:pot_oge_cornwall}
to lift some degeneracies between the states. This is not the case if we implement transverse
gluons by means of the Jacob and Wick formalism, then automatically the degeneracies are
lifted.

This improvement was carried out in a work based on the Coulomb
gauge Hamiltonian where a relativistic kinetic energy was
used.\cite{Szczepaniak:1995cw} In this model, gluons are linked by
an adjoint string. The adjoint string tension $\sigma_A=(9/4)\sigma$
is expressed in terms of the well-known fundamental string tension
for mesons $\sigma$ through the Casimir scaling hypothesis supported
by lattice calculations.\cite{Bali:2000un} Using typical values for
the parameters, $\sigma=0.18$~GeV$^2$ for the fundamental string
tension (extracted from mesons Regge trajectory) and $\alpha_S=0.4$
for the strong coupling,  this model encodes the essential features
of glueballs.

The spectrum of the Coulomb gauge Hamiltonian was in good agreement
with lattice QCD. Moreover, the singlet $2^{-+}$ and $2^{++}$ are
degenerate as in the Barnes'~model, a characteristic of the helicity
formalism. The authors found a difference between the scalar and
pseudoscalar glueballs. This splitting, about 250~MeV was
nevertheless not as strong as in lattice QCD (800~MeV).

Recently,  this problem was revisited keeping the basic ingredients
needed for obtaining an acceptable pure gauge spectrum compatible
with lattice results, {\it i.e.} semi-relativistic energy and the
helicity formalism for two transverse gluons.\cite{Mathieu:2008bf} A
simple Cornell potential was used but an instanton induced force was
added and with it the splitting between the scalar and pseudoscalar
glueballs was reproduced. There are arguments favoring an attractive
(repulsive) interaction induced by instantons in the scalar
(pseudoscalar) channel of glueballs.\cite{Schafer:1994fd}

In all the constituent models for glueballs, the confining
interaction follows from phenomenological considerations (breakable
strings or linear potentials). But it is also possible to derive an
effective Hamiltonian for bound states from the QCD Lagrangian.
Kaidalov and Simonov used the field correlator method to extract a
relativistic Hamiltonian which describes an adjoint string with
gluons at its ends.\cite{Kaidalov:1999yd} The introduction of
auxiliary fields (or einbein field) $\mu$ and $\nu$ leads to

\begin{equation}\label{eq:Hstring_simonov}
H_0 = \frac{p^2_r}{\mu}+\mu +
\frac{L(L+1)}{r^2\left[\mu+\int_0^1(\beta-\frac{1}{2})^2\nu
d\beta\right]} + \int_0^1\frac{\sigma_A^2 d\beta}{2\nu}r^2 +
\frac{1}{2}\int_0^1\nu d\beta.
\end{equation}
In Eq. \eqref{eq:Hstring_simonov}, $ \mu$ is introduced to get rid
of the square root and $\langle\mu\rangle = \langle\sqrt{\bm
p^2}\rangle$ represents the effective gluon energy.
$\langle\nu\rangle$ is the energy stored in the rotating string. The
einbeins are eliminated using the equations of motion, $\delta_\mu
H_0=\delta_\nu H_0=0$.  For $L=0$, we find $\nu=\sigma_A r$, and the
replacement of the auxiliary field leads to

\begin{equation}\label{}
    H_0 = 2\sqrt{\bm p^2}+\sigma_A r.
\end{equation}
For $L\neq0$ it is not possible to eliminate $\nu$ analytically and
the optimization $\delta_\nu H_0=0$ should be done numerically.

In this approach, gluons are massless but gain an effective mass
$\mu_0=\langle\mu\rangle\sim 0.5$-1~GeV. The expectation value is
taken on an eigenstate of $H_0$ therefore $\mu_0$ is
state-dependent. The spin splitting operators are corrections of
order $\mu_0^{-2}$ and are computed perturbatively. These
corrections contain the conventional structure arising from the
one-gluon-exchange and, in addition, a spin-orbit term coming from
the Thomas precession,

\begin{equation}\label{}
\begin{split}
  \Delta H =& -\frac{\lambda}{r}+\frac{\lambda}{r^3}\frac{\bm L\cdot\bm S}{\mu_0^2} +
  \frac{8\pi\lambda}{3\mu_0^2}\delta(\bm r)\bm S_1\cdot\bm S_2 \\
    &+
  \frac{\lambda}{r^3\mu_0^2}\left(3(\bm S_1\cdot\hat{\bm r})(\bm S_2\cdot\hat{\bm r})-\bm S_1\cdot\bm
  S_2\right)\\
  &-\frac{\sigma_A}{r}\frac{\bm L\cdot\bm S}{2\mu_0^2}
\end{split}
\end{equation}
with $\lambda=3\alpha_s$.

\begin{figure}[htb]
\epsfig{file=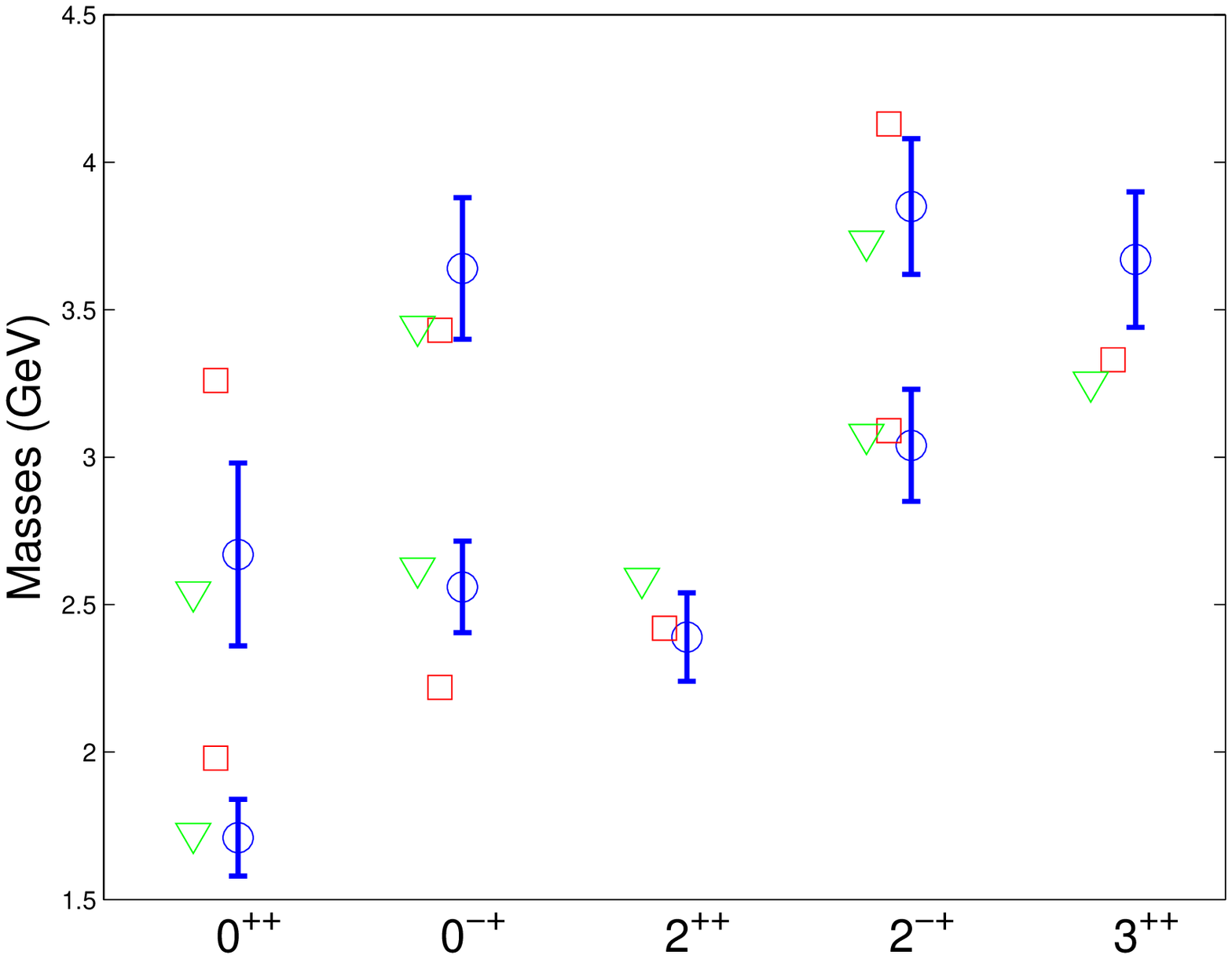,width=0.48\linewidth}
\epsfig{file=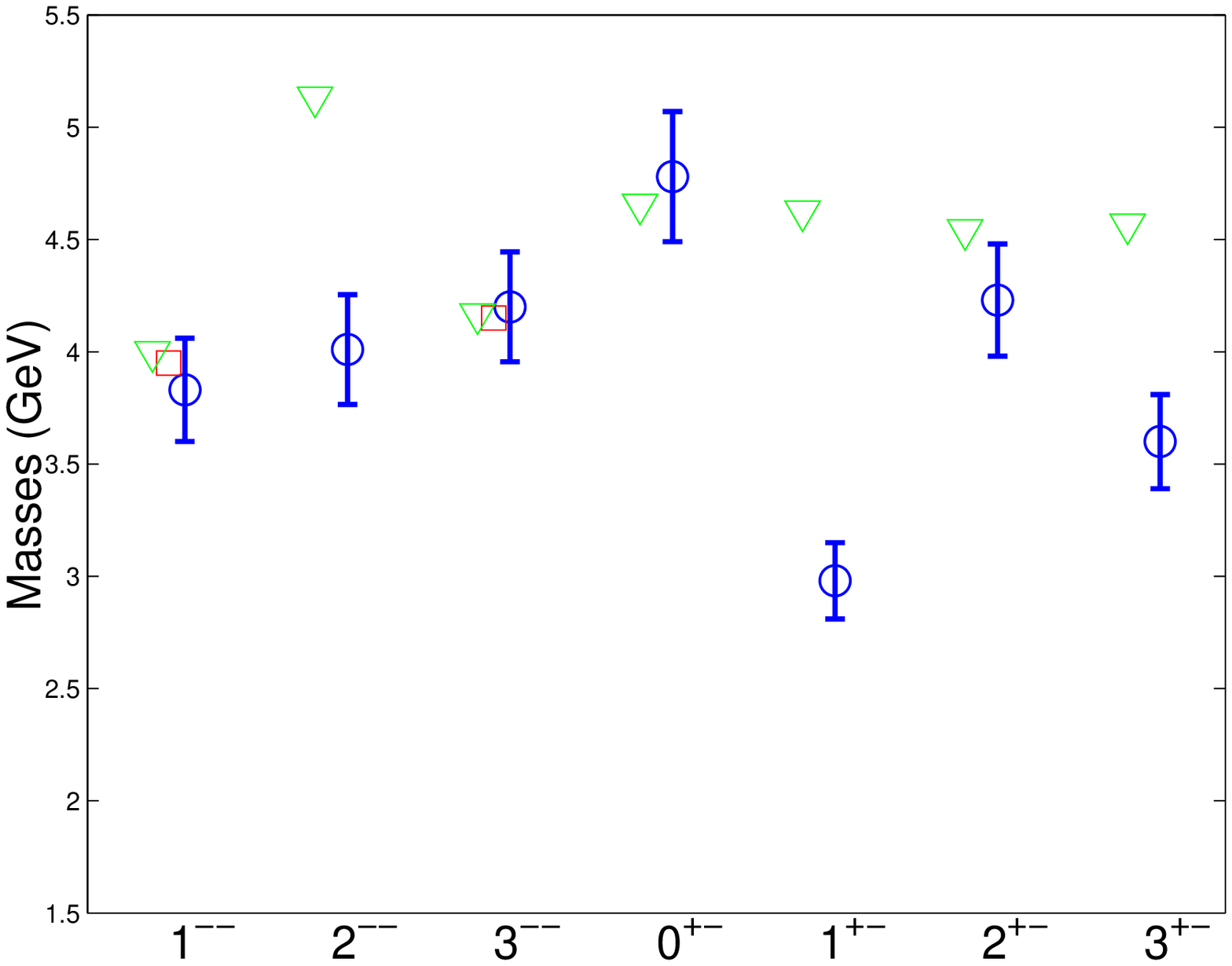,width=0.48\linewidth} \caption{Comparison between lattice
results~17~(circles) and two-gluon glueballs (left)  from Ref.~47~(squares) and
Ref.~45~(triangles) and three-gluon glueballs (right) from Ref.~55~(squares) and
Ref.~57~(triangles).\label{fig:2gluons}}

\end{figure}

The numerical results for the two-gluon glueballs in all the models
discussed are displayed in Table~\ref{tab:gg} and
Fig.~\ref{fig:2gluons}

\begin{table}[ht]
\caption{Two-gluon glueballs spectra in different models. In brackets, the ratios with respect to the scalar glueball. Masses are in GeV.}
\label{tab:gg} %\setlength{\extrarowheight}{4pt} \centering
\begin{tabular}{c|ccccc|cc}
\hline\hline $J^{PC}$& ref.~\cite{Barnes:1981ac} & ref.~\cite{Szczepaniak:1995cw} &
ref.~\cite{Mathieu:2008bf} & ref.~\cite{Kaidalov:2005kz} & &
ref.~\cite{Morningstar:1999rf,Chen:2005mg} & ref.~\cite{Meyer:2004jc}
\\
\hline
$0^{++}$& (1.00)& 1.98(1.00)& 1.72(1.00) & 1.41(1.00) && 1.71(1.00) & 1.48(1.00)\\
        & (1.40)& 3.26(1.65)& 2.54(1.48) & 2.41(1.71) && 2.67(1.56) &  2.76(1.89) \\
$0^{-+}$& (1.00)& 2.22(1.12)& 2.62(1.52) & 2.28(1.62) && 2.56(1.50) &  2.25(1.54) \\
        & (1.40)& 3.43(1.73)& 3.44(2.00) & 3.35(2.38) && 3.64(2.13) &  3.37(2.31) \\
$2^{++}$& (1.16)& 2.42(1.22)& 2.59(1.51) & 2.30(1.63) && 2.39(1.40) &  2.15(1.47) \\
        & (1.35)& 3.11(1.57)& 3.08(1.79) & 3.32(2.35) & &           & 2.88(1.97)\\
$2^{-+}$& (1.35)& 3.09(1.56)& 3.08(1.79) & 2.70(1.91) & &3.04(1.78) & 2.78(1.90)  \\
        & (1.68)& 4.13(2.09)& 3.73(2.17) & 3.73(2.65) && 3.89(2.27) & 3.48(2.38)  \\
$3^{++}$& (1.42)& 3.33(1.68)& 3.25(1.89) &      && 3.67(2.15) & 3.39(2.32)  \\
$4^{++}$& (1.63)& 3.99(2.02)& 3.77(2.19) &      &&  & 3.64(2.49)  \\
        & (1.71)& 4.28(2.16)& 3.96(2.30) &      &&  &   \\
$4^{-+}$& (1.71)& 4.27(2.16)& 3.96(2.30) &  &&  &   \\
$5^{++}$&       &           & 4.21(2.45) &  &&  & \\
$6^{++}$&       &           & 4.60(2.67) &  &&  & 4.36(2.98)\\
\hline  \hline
\end{tabular}
\end{table}

The low-lying positive $C$-parity glueball states, seem to favor a
constituent picture with two gluons interacting via a linear
potential, {\it i.e.} linked by a string. A relativistic approach with two
transverse gluons leads to a spectrum in good agreement with the
lattice hierarchy.

This stringy picture leads to a Regge trajectory

\begin{equation}\label{}
J \sim \frac{1}{2\pi\sigma}M^2,
\end{equation}
well-known in the meson sector. The experimental slope for mesons
$1/(2\pi\sigma)\approx0.9$~GeV$^{-2}$ corresponds to the typical
value for a fundamental string tension
$\sigma\approx0.18$~GeV$^{2}$. When we extend this picture to
glueballs, the slope rises to $0.4$~GeV$^{-2}$ if the Casimir
scaling hypothesis is used. If one argues that even spin positive
$C$-parity glueballs lie on the pomeron trajectory,  a problem
arises. Indeed, the experimental soft pomeron slope is $\alpha^\prime\approx0.25$ GeV$^{-2}$.
Since the pomeron trajectory
carries physical particles, a first solution is that glueballs
probably mix with quark states. Let us note before finishing this
discussion that to describe negative $C$-parity states in
constituent models one needs at least three gluons.

A very different approach to the interpretation of the spectrum is
in terms of a closed flux tubes. This picture was introduced  by
Isgur and Paton.\cite{Isgur:1984bm} This model is composed by a
closed loop of fundamental flux with no constituents gluons at all.
Such a loop has phononic and orbital degrees of excitation. These
two modes lead to two Regge behaviors at large spin,

\begin{align}\label{eq:regge_closed_TF}
\text{phononic :  } J \sim& \frac{1}{8\pi\sigma}M^2 ,& \text{orbital
:  } J \sim& \frac{3\sqrt{3}}{32\pi\sigma}M^2.
\end{align}
In either case one obtains a slope $0.2$-$0.3$~GeV$^{-2}$ which is in the right range for the
pomeron [see Eq.~\eqref{eq:pomeron}].\cite{Meyer:2004jc} Another interesting feature of the
closed flux tube model is the appearance of low-lying odd spin $PC=+-$ states in the spectrum.
In their original paper, Isgur and Paton discussed masses for some low-lying pure glue states
(see Fig. \ref{tab:isgur}).\cite{Isgur:1984bm}. These authors argued that the true values of
their parameters were not known and they simply chose them to fit expected results. The
splittings are not sensitive to their choice of parameters but their absolute values are.

\begin{table}[ht]
\caption{Low lying glueball states in the flux tube model.}
\label{tab:isgur}
\begin{center}
\begin{tabular}{l c}
\hline\hline \hskip 1.0 cm$J^{PC}$& Mass (GeV) \\
\hline
$0^{++}$& 1.52\\
$1^{+-}$& 2.25 \\
$0^{++}$& 2.75 \\
$0^{++},0^{+-},0^{-+},0^{--}$& 2.79 \\
$2^{++}$&  2.84\\
$2^{++},2^{++},2^{++},2^{++} $& 2.84  \\
$1^{+-}$& 3.25 \\
$3^{+-}$&  3.35    \\

\hline  \hline
\end{tabular}
\end{center}
\end{table}

\subsection{Three-gluon glueballs}

A complete investigation of the glueball spectrum in constituent
models has to include three-gluon glueballs. Indeed, in this
approach negative $C$-parity glueballs involve at least three
constituents. There are two color wave functions, totally symmetric
or totally antisymmetric, coupling three adjoint representations
into a singlet. They do not mix and we are only interested in the
symmetric one, $d_{abc}A^a_\mu A^b_\nu A^c_\rho$, namely the $C=-$
states.

Soni with Hou extended his paper with Cornwall to three massive
gluons.\cite{Hou:1982dy} Their potential is a generalization of the
previous one,

\begin{equation}\label{}
V_{ggg}=\sum_{i<j}  \frac{1}{2} V_C(r_{ij})+V_{ggg}^{oge}(r_{ij}),
\end{equation}
with $V_{C}$ the confining potential of the breakable string of Eq.~\eqref{eq:pot_sat}. A
factor one-half is added because one needs to remove three (and not six) gluons from the
vacuum to screen the three gluons that are originally there in the glueball. The short-range
potential is not the OGE potential Eq.~\eqref{eq:pot_oge_cornwall} with a different color
factor, because one should also take into account the annihilation diagram not present in
two-gluon glueballs, but it involves the same structures (spin-spin, tensor and spin-orbit
interactions). The nonrelativistic kinetic energy and the screened confining potential lead
to a spectrum which is too low. The low-lying states are nearly degenerate with a mass $4.8$
times the constituent mass $m$ of the gluons. By low-lying, the authors mean that every pair
of particles is an $S$-wave. Symmetry considerations on the total wave function imply that the
low-lying states are the $0^{-+}, 1^{--}, 3^{--}$.

The $3^{--}$ is the lowest maximum spin state. One often argues that
such states with maximal spin lie on the odderon trajectory, the
counterpart to the pomeron trajectory in hadron-hadron scattering at
high energy. The linearity of the odderon trajectory was checked
using a Coulomb gauge Hamiltonian.\cite{LlanesEstrada:2005jf} Their
masses for the $1^{--},3^{--},5^{--},7^{--}$ are shown in
Table~\ref{tab:odderon}.

\begin{table}[ht]
\begin{center}
\caption{\label{tab:odderon} Odderon quantum numbers and  masses in
MeV.}
\begin{tabular}{ccccccc}
\hline\hline $J^{PC}$& Conf. & $\sigma$ (GeV$^2$)&  $1^{--}$ &  $3^{--}$ & $5^{--}$ & $7^{--}$
\\ \hline
$S$  &&       & 1  &  3 & 3 & 3 \\
$L$  &&       & 0  & 0 & 2 & 4 \\
\hline
Llanes-Estrada\cite{LlanesEstrada:2005jf}& $\Delta$ & 0.18 &  3950 & 4150 & 5050 & 5900 \\
Simonov\cite{Kaidalov:1999yd}& $\Delta$ & 0.238 & 3490 &  4030 & & \\
Simonov\cite{Kaidalov:2005kz}& $\Delta$ & 0.18 & 3020 &  3490 & 4180 & 4960\\
& $Y$ & 0.18 & 3320 &  3830 & 4590& 5250\\
Mathieu\cite{Mathieu:2006bp,Mathieu:2008pb} & $Y$ & 0.21 & 3999 & 4167 & 5263 &  \\
Lattice\cite{Chen:2005mg}&   & 0.1939 &  3830  & 4200 & &  \\
Lattice\cite{Meyer:2004jc}& & 0.1939 & 3100 & 4150 & &  \\
\hline\hline
\end{tabular}\end{center}
\end{table}

In their Coulomb gauge approach to three-gluon glueballs, these
authors chose a spin-independent potential which is a sum of
two-body Cornell ones and a $\Delta$-shape for confinement. Another
Ansatz for the confinement, the $Y$-shape, is sometimes preferred.
The $Y$-shape is the generalization of confinement in baryons where
every quark provides a flux tube. These flux tubes coming from
gluons meet in a point where the total energy (or the length since
the energy density is constant) is minimal. It is worth mentioning
that, under the Casimir scaling hypothesis, a simple application of
the triangular inequalities shows that the $\Delta$-shape is
energetically more favorable than the $Y$-shape.\cite{Mathieu:2005wc}
This demonstration was recently confirmed by a lattice study of the three-gluon potential.\cite{Cardoso:2008sb}

Kaidalov and Simonov investigated both Ans\"atze for confinement
with a nonrelativistic Hamiltonian\cite{Kaidalov:2005kz}
\begin{equation}\label{eq:H3g_simonov}
H_{ggg}=\frac{\bm p_1^2+\bm p_2^2 + \bm p_3^3}{2\mu}+\frac{3\mu}{2} +
V_{\Delta,Y}(\bm r_1,\bm r_2,\bm r_3),
\end{equation}
with
\begin{align}\label{}
    V_{\Delta}(\bm r_1,\bm r_2,\bm r_3)&=\sigma\sum_{i<j}|\bm r_i-\bm r_j|,  &
    V_{Y}(\bm r_1,\bm r_2,\bm r_3)&=\frac{9}{4}\sigma\sum_{i=1}^3|\bm r_i-\bm R_{Y}|.
\end{align}
They found the eigenvalues of this operators by using the
hyperspherical formalism. The spin-average masses of this
Hamiltonian for the odderon states (odd)$^{--}$ are displayed in
Table~\ref{tab:odderon}. Clearly, the $Y$-shape potential leads to a
higher spectrum than the $\Delta$-shape one.

\begin{table}[ht]\begin{center}
\caption{Glueball spectrum in the $PC=+-$ sector.}
\begin{tabular}{llll}
\hline \hline
 $J^{PC}$ & Lattice\cite{Chen:2005mg} &  Mathieu\cite{Mathieu:2008pb} & Simonov\cite{Kaidalov:2005kz} \\
\hline
$0^{+-}$ & 4780  &  4656  & 4090\\
$1^{+-}$ & 2980  &   4626 & 4090\\
$2^{+-}$ & 4230  &   4542 & 4090\\
$3^{+-}$ & 3600 &    4568 & 4090 \\
$5^{+-}$ & 4110 \cite{Meyer:2004jc}  & 5317 &   \\
\hline \hline
\end{tabular}\end{center}
\end{table}

As we saw in the case of two-gluon glueballs, a nonrelativistic
description of such states is not appropriate. However, in the
negative $C$-parity sector, the lack of lattice results for high
spin states does no allow us to draw any definitive conclusions.
Also, the Regge trajectory of the odderon is not well understood
yet. New theoretical and experimental research on these topics would
be helpful for understanding these glueballs.

On the other hand, lattice QCD exhibits an interesting spectrum in
the $PC=+-$ sector. There is nearly 2 GeV between the lowest
$1^{+-}$ state and the highest $0^{+-}$. According to the $LS$
decomposition in constituent models, all these states
($0^{+-},1^{+-},2^{+-},3^{+-}$) should be $L=1$ and degenerate. This
feature appears clearly in Simonov and Kaidalov's
paper.\cite{Kaidalov:2005kz}

Using a generalization of the Hamiltonian Eq.~\eqref{eq:H3g_simonov}
but with a semi-relativistic kinetic energy, Mathieu {\it et. al}
found also a degeneracy between the $J^{+-}$ states in disagreement
with the lattice results.\cite{Mathieu:2008pb} Their Hamiltonian
involves a $Y$-junction at the center of mass for the confinement
and an OGE potential for the short-range part

\begin{equation}\label{}
H = \sum_{i=1}^{3}\sqrt{\bm p^2_i}+a \sum_{i=1}^{3} |\bm r_i- \bm
R_0|+ V_{\text{OGE}}.
\end{equation}

Their study is an extension of a first one where only $L=0$ states were
considered.\cite{Mathieu:2006bp} The parameters for this three-gluon glueball model  were
fitted on the two-gluon glueballs of a previous work.\cite{Brau:2004xw} Previously, the
$2^{--}$ had been computed with the wrong wave function.\cite{LlanesEstrada:2005jf} In this
study, the right $2^{--}$ state was obtained. It was noted that this result disagrees with
that of lattice QCD.\cite{Mathieu:2006bp} In a model with longitudinal gluons, the $2^{--}$
cannot lie between the $1^{--}$ and the $3^{--}$ as lattice calculations have shown (cf.
Fig~\ref{fig:2gluons}). This was a hint that a description with the nonrelativistic
decomposition $\bm J=\bm L+\bm S$ should be inadequate to handle the pure gauge spectrum. In
the $PC=+-$ sector, all states within this approach lie in the same range and they contradict
the lattice results. Hence, we conclude that a description of many-gluon states with a
$LS$-basis is not appropriate. A solution could be to implement the helicity formalism for
three particle states for three transverse gluons. One hopes that then it would be possible to
reproduce the correct hierarchy. Indeed, the lowest states with three longitudinal gluons are
the $1^{--}$ and $3^{--}$ while in lattice QCD the lowest negative $C$-parity are the $1^{+-}$
and $3^{+-}$.

\subsection{AdS/QCD}

An alternative approach to strong interactions is based on the idea that they have a
description in terms of strings. A remarkable step in this direction was given by Maldacena
proposing the equivalence between conformal fields and string theory in anti-de Sitter
spacetime (AdS/CFT correspondence).\cite{Maldacena:1997re} In particular glueball operators of
the conformal gauge theory defined on the AdS boundary are in correspondence with the string
dilaton field. The description of the strong interactions based on this correspondence
requires the breaking of conformal invariance, which can be done in different ways. The
spectrum of the glueball has been calculated in two of these models for breaking. The first
introduces a Schwarzschild black hole in AdS to break scale invariance.\cite{Witten:1998zw}
The corresponding supergravity equations do not admit closed form solutions and the spectrum
has been calculated using approximate methods (see Table
\ref{tab:ads}).\cite{Csaki:1998qr,Brower:2000rp} The second possibility is to imitate the
Randall-Sundrum model by considering two slices of AdS sticked together assuming that there is
a bulk/boundary correspondence between the glueballs and the dilaton.\cite{BoschiFilho:2002vd}
Imposing boundary conditions the string dilaton acquires discrete modes and these modes become
the spectrum of glueballs (see Table \ref{tab:ads}).

\begin{table}
\begin{center}
\caption{Masses of glueball states $J^{PC}\,$ with even $J$ expressed in GeV, estimated using
the sliced $AdS_5\times S^5$ space with Dirichlet (Neumann) boundary conditions.$^{63}$ The
mass of $0^{++}$ is an input from lattice results.$^{17,64}$ \label{tab:ads}} \vskip 0.5cm
\begin{tabular}{ | c | c | c | c |}
\hline Dirichlet $\,\,$
 & $\,\,$ lightest $\,\,$   &
$1^{st}$ excited $\,\,$ & $2^{nd}$ excited \\
glueballs $\,\,$     &   state & state & state \\
 \hline
 $0^{++}$ & 1.63  &  2.67 & 3.69 \\
 $2^{++}$ & 2.41    & 3.51  & 4.56  \\
 $4^{++}$ & 3.15  & 4.31  & 5.40  \\
 $6^{++}$ & 3.88  & 5.85  & 6.21 \\
 $8^{++}$ & 4.59 & 5.85  & 7.00 \\
 $10^{++}$ & 5.30 & 6.60   & 7.77 \\
\hline
\end{tabular}
 \vskip 1.0cm
\begin{tabular}{ | c | c | c | c |}
\hline Neumann $\,\,$
 & $\,\,$ lightest $\,\,$   &
$1^{st}$ excited $\,\,$ & $2^{nd}$ excited \\
glueballs $\,\,$     &   state & state & state \\
 \hline
 $0^{++}$ & 1.63  & 2.98  & 4.33  \\
 $2^{++}$ & 2.54    & 4.06 & 5.47  \\
 $4^{++}$ & 3.45 & 5.09  & 6.56 \\
 $6^{++}$ & 4.34  & 6.09 & 7.62 \\
 $8^{++}$ & 5.23 & 7.08 & 8.66 \\
 $10^{++}$ & 6.12 & 8.05  & 9.68 \\
\hline
\end{tabular}
\end{center}
\end{table}

Let us for the purposes of illustration, describe the  analysis of
the complete glueball spectrum calculation for the $AdS^7$ black
hole supergravity dual of $QCD_4$ in strong coupling limit: $g^2 N
\rightarrow \infty$.\cite{Brower:2000rp} Despite the expected limitations of
a leading order strong coupling approximation, the pattern of
spins, parities and mass inequalities bare a striking resemblance
to the known $QCD_4$ glueball spectrum as determined by lattice
simulations at weak coupling.

\begin{center}
\begin{figure}[htb]
\begin{minipage}{6.0cm}
\vspace{-0.15cm}
\begin{center}
\epsfig{file=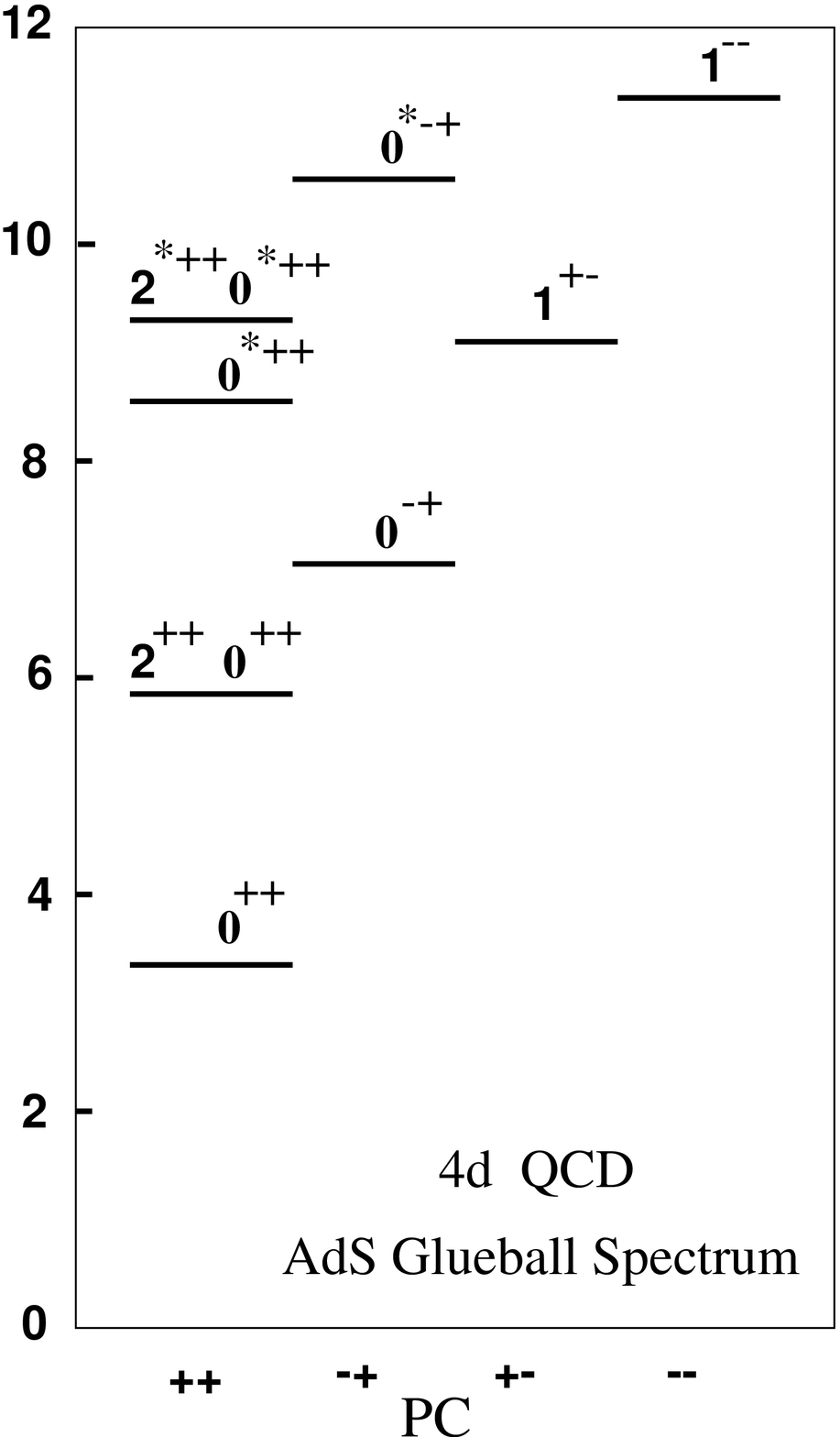,width=0.88\linewidth,height=1.1\linewidth}
\end{center}
\end{minipage}
\begin{minipage}{6.0cm}
\begin{center}
\epsfig{file=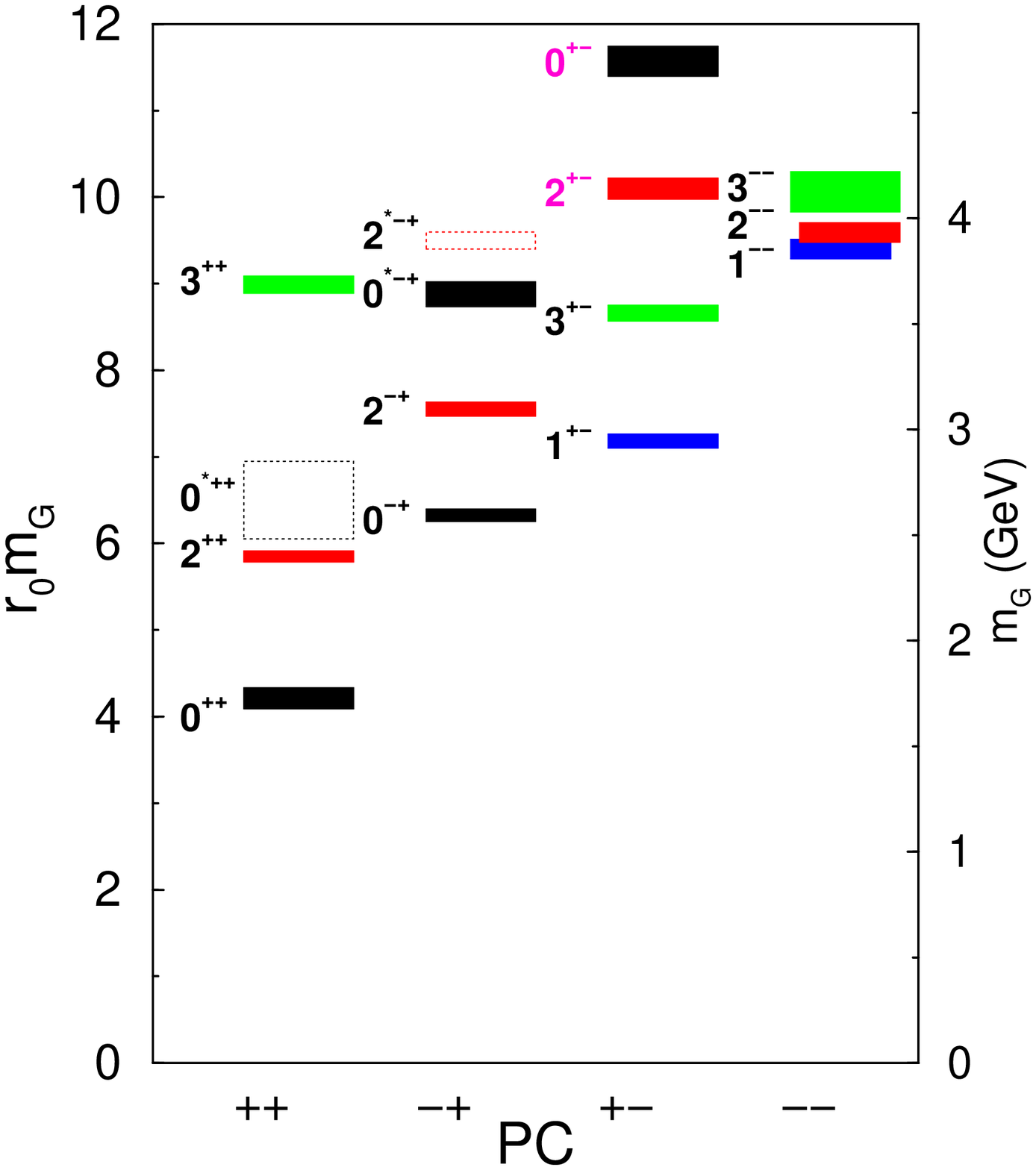,width=1.05\linewidth,height=1.16\linewidth}
\end{center}
\end{minipage}
\caption{The AdS glueball spectrum$^{61}$ for $QCD_4$ in strong
        coupling (left) compared with the lattice spectrum$^{17}$ for pure
        SU(3) QCD (right). The AdS cut-off scale is adjusted to set the
        lowest $2^{++}$ tensor state to the lattice results in units of
        the hadronic scale $1/r_0 = 410$ Mev.\label{fig:ads}}
\end{figure}
\end{center}

To approach $QCD_4$ one
begins with M theory on ${\bf AdS^7 \times S^4}$. One compactifies
the ``eleventh'' dimension (on a circle of radius $R_1$) to reduce
the theory to type IIA string theory and then following the
suggestion of Witten raise the ``temperature'', $\beta^{-1}$, with a
second compact radius $R_2$ in a direction $\tau$, with $\beta =
2\pi R_2$. On the second ``thermal'' circle, the fermionic modes
have anti-periodic boundary conditions breaking conformal and all
SUSY symmetries.  This lifts the fermionic masses and also the
scalar masses, through quantum corrections. The 't Hooft coupling is
$g^2 N =2\pi g_s N l_s/R_2$, in terms of the closed string coupling,
$g_s$ and the string length, $l_s$ .  Therefore, in the scaling
limit, $g^2 N \rightarrow 0$, if all goes as conjectured, there
should be a fixed point mapping type IIA string theory onto $SU(N)$
pure Yang-Mills theory.

One considers the strong coupling limit at large N, where the string
theory becomes classical gravity in the $AdS^7$ black hole metric,

\begin{equation}
 ds^2 = (r^2-{1\over r^4}) d\tau^2 + r^2 \sum_{i=1,2,3,4,11} d
x_i^2 + (r^2-{1\over r^4})^{-1} d r^2 +{1\over 4} d\Omega_4^2 \; ,
\end{equation}
with radius of curvature, $R^3_{AdS} = 8 \pi g_s N l_s^3$.  The
dimensionful parameters have been removed in the metric by a
normalization setting $R_{AdS}= 1$ and $\beta=2\pi/3$.

The strong coupling glueball calculation consists of finding the
normal modes for the bosonic components of the supergraviton
multiplet in the ${\bf AdS^7 \times S^4}$ black hole background.  We
are only interested in excitations that lie in the superselection
sector for $QCD_4$. Imposing this restriction and exploiting
symmetries of the background metric reduce the problem to  six
independent wave equations which have to be solved with the
appropriate boundary conditions.

There is a rather remarkable correspondence of the overall mass and
spin structure between  strong coupling glueball spectrum and the
lattice results at weak coupling for $QCD_4$, see
Fig.~\ref{fig:ads}. Apparently the spin structure of type IIA
supergravity does resemble the low mass glueball spin splitting. The
correspondence is sufficient to suggest that the Maldacena duality
conjecture may well be correct and that further efforts to go beyond
strong coupling are worthy of sustained effort.

The basic idea behind the AdS/CFT correspondence is that the low mass glueball spectrum can be
qualitatively understood in terms of local gluon interpolating operators of minimal dimension
an idea which is not new.\cite{FritzschMinkowski,Jaffe:1985qp} These operators are in rough
correspondence with all the low mass glueballs states, as computed in a constituent gluon or
bag models discussed previously. Thus the AdS/QCD if valid would justify the behavior of the
gluon  as a constituent. Let us note, however, that in constituent models, confinement is
ensured by the potential and in the bag model and AdS/QCD, by a boundary conditions.

\section{Glueballs and QCD Sum Rules}\label{sec:sumrules}

The QCD sum rule (SR) approach is one of the most widely used methods to obtain the
information about glueball
properties.\cite{Novikov:1979va,Shuryak:1982dp,Zhang:2003mr,Narison:2005wc} It is based on the
operator product expansion (OPE) of the correlator of two glueball interpolating currents
$J_G$
\begin{equation}
\Pi(Q^2)=i\int d^4x
 \ e^{iq\cdot x}\langle 0| T J_G(x)J_G(0)|0\rangle
\label{correlator}
\end{equation}
in the deeply-Euclidean domain $Q^2=-q^2\gg\Lambda_{QCD}^2$. The method takes into account
perturbative, as well as, nonperturbative gluonic contributions to this correlator. The
perturbative contributions arise by direct calculation of Feynman diagrams, while the non
perturbative contributions are associated, as we shall discuss, with vacuum expectations
values of the correlators, {\it i.e.} condensates, and sometimes with direct instanton
contributions.

The interpolating currents, consistent with the minimum number of gluon fields, used to study
low mass glueballs  are the field strength squared ($S=0^{++}$), the topological charge
density ($0^{-+}$) and the energy-momentum tensors ($T=2^{++}$): \footnote{In
refs.~\cite{Latorre:1987wt,Hao:2005hu} three-gluon interpolating currents were used for the
scalar and pseudoscalar glueballs.}
\begin{eqnarray}
J_S(x)&=&\alpha_sG_{\mu\nu}^a(x)G_{\mu\nu}^a(x),\label{pseudo}\\
J_P(x)&=&\alpha_sG_{\mu\nu}^a(x)\widetilde
G_{\mu\nu}^a(x),\label{scalar}\\
J_T^{\mu\nu}(x)&=-&G_{\mu\alpha}^a(x)G_{\nu\alpha}^a(x)+
\frac{g^{\mu\nu}}{4}G_{\beta\alpha}^a(x)G_{\beta\alpha}^a(x),\label{tensor}
\end{eqnarray}
where $\widetilde
G_{\mu\nu}^a(x)=(1/2)\epsilon_{\mu\nu\alpha\beta}G_{\alpha\beta}^a(x)$.

The dispersion relation
\begin{equation}
\Pi(Q^2)=\frac{(-Q^2)^n}{\pi}\int_0^{\infty}ds\ \frac{{\rm
Im}\Pi(s)}{s^n(s+Q^2)}+ a_0+a_1Q^2+...,  \label{dis}
\end{equation}
where $a_i$ are substraction constants, allows us to connect the
theoretical side of the SR (left hand side) with observed properties
of the glueballs introduced through ${\rm Im } \Pi (s)$ (right hand
side). Conventionally, for the phenomenological part of the SR, the
following  form for the spectral representation is used,

\begin{equation}{\rm Im}\Pi(s)^{\text{phen}}=\sum_i\pi f_{G_i}^2m_{G_i}^4\delta(s-m_{G_i}^2)+
\pi\theta(s-s_0)\;{\rm Im}\Pi(s)^{\text{theor}}, \label{spectral}
\end{equation}
which corresponds to a sum over narrow width resonances located at
$s_i=M_{G_i}$ plus a continuum at large energy $s>s_0$. In
Eq.~\eqref{spectral}, $f_{G_i}$ is the residue of the
$i^{\text{th}}$-glueball state defined by the following matrix
element,
\begin{equation}
\langle0|J_G(0)|G_i\rangle=m_{G_i}^2f_{G_i}. \label{res}
\end{equation}
Here, $m_{G_i}$ is $i^{\text{th}}$-glueball mass, $s_0$ is the continuum threshold and
$\Pi(s)^{\text{theor}} $ is the perturbative part of the correlator. This is, of course,  a
rather simplified  model for the physical spectral density. However, it has been shown that
such a model gives a rather good description of the mass spectrum for the ordinary hadronic
states. One may expect therefore, that such a model can be also used for the extracting of the
glueball masses.

Before carrying out the numerical analysis of the SR one usually
performs the Borel transformation on both sides of the sum rule
\begin{equation}
\hat{B}\Pi(Q^2)\equiv\lim_{n,Q^2\rightarrow\infty}\frac{(-1)^n}{n!}(Q^2)^{n+1}
\bigg(\frac{d}{dQ^2}\bigg)^n\Pi(Q^2)\bigg|_{Q^2/n=M^2={\rm fixed}}\
\nonumber
\end{equation}
where $M$ is called the Borel mass which represents the scale
$\tau=1/M$ in Euclidean time between the two currents in the correlator,
Eq.~\eqref{correlator}.\footnote{Some examples of Borel
transformations are
$\hat{B}\bigg(\frac{1}{p^2-\alpha}\bigg)^\beta=(-1)^\beta(M^2)^{1-\beta}
\frac{e^{-\alpha/M^2}}{(\beta-1)!}$ ,
$\hat{B}\bigg((p^2)^m\ln(-p^2)\bigg)=-m!(M^2)^{m+1}$ and
$\hat{B}\bigg((p^2)^m\bigg)=0$, ${\rm for}\ m\geq 0.$} The Borel
transformation of the sum rules  has two major advantages. The
first, it produces an enhancement of the glueball pole contribution
in the right hand side of the SR and, the second, it suppresses, in
the theoretical left hand side of the SR, the contribution from the
higher power corrections, which arise from high dimension
condensates.

The main point of the QCD SR philosophy is the implementation of the interaction of the high
virtual valence quark-gluon system with the soft vacuum quark and gluon fields, whose strength
is determined by the values of the vacuum condensates. The condensates carry very important
information about the long-range quark and gluon field fluctuations in the QCD vacuum. The
usual assumption is that one can calculate the hard part within perturbative QCD and the soft
part can be parameterized in terms of condensates. It has been demonstrated that, within the
OPE approach, in general, the contribution of only a few low dimension condensates is
sufficient to describe rather well the properties of the ground hadronic
states.\cite{Reinders:1984sr} However, it was found, many years ago, that in some specific
channels, which include the quark-gluon subsystem in a spin zero state, the standard OPE
expansion does not work and one needs to incorporate more precise information about the
structure of the short-range gluonic fluctuations in the QCD
vacuum.\cite{Shuryak:1982dp,Geshkenbein:1979vb,Dorokhov:1989zw} The most promising candidate
for those short-range fluctuations are instantons (see ref.~\cite{Schafer:1996wv}). The
instantons describe tunneling processes which rearrange the QCD vacuum topology in localized
space-time regions small enough to affect the $x$-dependence of the correlators over distances
$\left| x\right| \ll\Lambda_{\text{QCD}}^{-1}$. The average size of instantons in the QCD
vacuum is small $\rho_c\approx 0.3$ fm. Therefore, the energy scale related to instanton
effects is rather large, $m_I=1/\rho_c\approx 600$ MeV$\approx 3\Lambda_{\text{QCD}}$. Due to
such large scale, instantons influence strongly not only  the dynamics of the low mass
pseudoscalar and scalar mesons, but also the properties of the baryon octet and multiquark
states.\cite{Lee:2004dp,Lee:2005ny,Lee:2006vk} It has been shown that specific instanton
induced quark-quark, quark-gluon and gluon-gluon interactions arising from intermediate
distances between hadron constituents can be responsible for various observed features in
hadron spectroscopy and hadron
reactions.\cite{Schafer:1996wv,Dorokhov:1993ym,Diakonov:2002fq,Kochelev:2005xn} The new
ingredient within QCD SRs related to instantons, the so-called direct instanton contribution,
provides exponential terms, $\sim \exp({-2Q\rho_c})$ in the expansion of the correlator, in
addition to the power terms, $\sim 1/Q^n$, arising from the standard OPE.

\begin{figure}[h]
\centerline{\epsfig{file=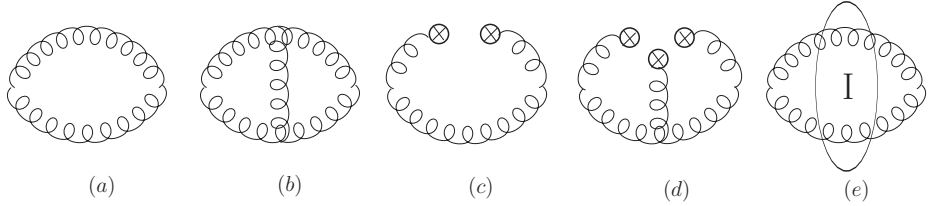,width=2.7cm,angle=90}}
 \caption{ The diagrams  (a) and (b) represent pQCD contributions,
 diagram (c) represents the contribution
arising from the gluon condensate, diagram (d) is the contribution from three-gluon condensate
and diagram (e) is the direct instanton contribution. } \label{ope}
\end{figure}

Finally, the theoretical part of the SR includes the following terms (see Fig.~\ref{ope})
\begin{equation}
\Pi^{\text{theor}}(Q^2)=\Pi^{\text{pert}}(Q^2)+\Pi^{\text{cond}}(Q^2)+\Pi^{\text{inst}}(Q^2).
\label{theor}
\end{equation}
The first two terms  in Eq.~\eqref{theor} are calculated using
Feynman rules giving for the perturbative part for the correlator of the
gluonic field strength\cite{Hao:2005hu}
\begin{equation}
\langle
TG_{\mu\nu}^a(x)G_{\alpha\beta}^b(0)\rangle_{\text{pert}}=-i\delta^{ab}
\int\frac{d^4p}{(2\pi)^4)}\Gamma_{\mu\nu\alpha\beta}(p)e^{-ipx},
\label{pert}
\end{equation}
where
\begin{equation}
\Gamma_{\mu\nu\alpha\beta}(p)=p_\mu p_\nu g_{\alpha\beta}+p_\nu
p_\beta g_{\mu\alpha}- p_\mu p_\beta g_{\nu\alpha}-p_\nu p_\alpha
g_{\mu\beta}. \nonumber
\end{equation}
The term associated to the vacuum condensates is evaluated in the
so-called fixed point gauge. In this gauge the soft vacuum gluonic
field can be represented by the field strength
\begin{equation}
{A_\mu^a(x)}_{\text{cond}}\approx\frac{x^\nu}{2}{G^a_{\nu\mu}}_{\text{cond}}(0). \label{fix}
\end{equation}
Using that the QCD vacuum is colorless and Lorenz invariant, and
Eq.~\eqref{fix}, one obtains the contribution of the vacuum
condensates to the OPE for the gluonic channels. It turns out
proportional to the value of gluon condensates,
\begin{equation}
\langle\alpha_sG^2\rangle=\langle\alpha_sG^a_{\mu\nu}G^a_{\mu\nu}\rangle,
\quad \langle g_sG^3\rangle=\langle g_sf^{abc}
G^a_{\mu\nu}G^b_{\nu\rho}G^c_{\rho\mu}\rangle.
\end{equation}
Unfortunately, the values of the gluonic condensates are not well known. Even the lowest
dimension gluon condensate $\langle\alpha_sG^2\rangle\approx 0.035$-0.075 GeV$^4$ is fixed by
considering the SRs for quark systems and has large uncertainties (see discussion in
ref.~\cite{Ioffe:2005ym}). Other gluonic condensates have been estimated using  models for the
QCD vacuum. For example, the tree gluon condensate within single-instanton approximation is
quite small $\langle g_sG^3\rangle\approx 0.27$~GeV$^2$ $\langle\alpha_sG^2\rangle$. Lattice
calculations are also uncertain. For the lowest dimension condensate the quenched value,
$0.14\pm .02$~GeV$^4$, is about one order of magnitude bigger than the unquenched one,
$0.022\pm.005$~GeV$^4$.\cite{D'Elia:1997ne} The uncertainties in the condensate values lead to
effects of the order of at least $20\%$ in the extracted masses for glueballs.

The direct instanton contribution is calculated by going to the Euclidian space-time and by
introducing in the correlator~\eqref{correlator} the instanton field strength, which in the
regular gauge has the following form
\begin{equation}
G_{\mu\nu}(x,x_0)=-\frac{\eta_{a\mu\nu}\rho^2}{g_s((x-x_0)^2+\rho^2)^2},
\label{inst}
\end{equation}
where $x_0$ is the position of center of instanton and
$\eta_{a\mu\nu}$ are the numerical t'Hooft symbols. To get the final
result one should also integrate over instanton position, size and
orientation.

Let us study in some detail the scalar glueball groundstate, $J^{PC} = 0^{++}$, which would be
an ideal state to find since it has many fundamental
connotations.\cite{Novikov:1979va,Migdal:1982jp,Ellis:1984jv,West:1996du,Vento:2004xx} In
gluodynamics, as we have seen, the situation that arises from lattice calculations is clear
and the masses of the scalar glueballs are large $m > 1$~GeV. However, when sea quarks are
considered no firm conclusion about the scalar spectrum can be drawn. The theoretical
calculations based on QCD SRs and/or low energy theorems lead to contradictory results. Its
properties, {\it i.e.}, mass, decay channels and widths still differ among the various
approaches. Dominguez and Paver \cite{Dominguez:1986td}, Bordes, Pe\~narrocha and
Gim\'enez \cite{Bordes:1989kc}, and Kisslinger and Johnson \cite{Kisslinger:2001pk}, obtain,
using low energy theorems and/or SR calculations with (or without) instanton contributions, a
low lying and narrow scalar glueball (mass $<$~700 MeV and $\Gamma_{\pi\pi} <$~100 MeV).
Narison and collaborators \cite{Narison:2002gv,Dosch:2002hc}, using two (substracted and
unsubstracted) SRs, get a broader (200-800~MeV), heavier (700-1000~MeV) scalar glueball in
this range, whose properties imply a strong violation of the Okubo-Zweig-Iizuka's rule. In a
recent SR calculation, Forkel \cite{Forkel:2003mk}, obtains the scalar glueball at
1250~$\pm$~200~MeV. However, he has some strength at lower masses which he is not able to
ascribe to a resonance in the fits.

Having made these comments let us now see how a calculation proceeds and we let the details of
the various approaches just described for the reader to make his own opinion. The calculation
we describe is in line with the work of Forkel.\cite{Forkel:2003mk}

In this case the perturbative contribution arising from Fig.~\ref{ope}(a) and (b) is
proportional to a high power of the $Q$
\begin{equation}
\Pi^{\text{pert}}(Q^2)=
(\frac{\alpha_s}{\pi})^2Q^4\log(Q^2/\mu^2)(A_1+\frac{\alpha_s}{\pi}A_2),
\end{equation}
where $A_i$ are some numbers. Therefore, one needs to apply three substractions to have
convergence for the dispersion relation\cite{Pascual:1982bv}
\begin{equation}
\Pi^{theor}(Q^2)=\Pi(0)+Q^2\Pi^{\prime}(0)+\frac{1}{2}Q^4\Pi^{\prime\prime}(0)-
\frac{Q^6}{\pi}\int_{0}^\infty ds\frac{{\rm Im}\Pi(s)}{s^3(s+Q^2)}
\label{sub}
\end{equation}
The SRs for the following Borel transforms
\begin{equation}
B_\kappa(\tau)=\frac{1}{\tau}\hat{B}[(-1)^\kappa
Q^{2\kappa}\Pi(Q^2)]
\end{equation}
are
\begin{eqnarray}
B_{-1}&=&-\Pi(0)+\frac{1}{\pi}\int_{0}^{\infty}\frac{ds}{s}e^{-s\tau}{\rm Im}\Pi(s)\nonumber\\
B_{\kappa}&=&\frac{1}{\pi}\int_{0}^{\infty}dss^ke^{-s\tau}{\rm
Im}\Pi(s), \ \ \kappa>-1. \label{sr1}
\end{eqnarray}
After moving the contribution from continuum to the left hand side
(theoretical part) of the SR we have
\begin{eqnarray}
S_{-1}(\tau,s_0)&=&-\Pi(0)+\frac{1}{\pi}\int_{0}^{s_0}\frac{ds}{s}e^{-s\tau}{\rm Im}\Pi(s)\nonumber\\
S_{\kappa}(\tau,s_0)&=&\frac{1}{\pi}\int_{0}^{s_0}dss^ke^{-s\tau}{\rm
Im}\Pi(s), \ \ \kappa>-1. \label{sr11}
\end{eqnarray}
Finally the SRs become
\begin{eqnarray}
S_{-1}(\tau,s_0)&+&\Pi(0)=\sum_i f_{G_i}^2m_{G_i}^2e^{m_{G_i}^2\tau}\nonumber\\
S_{\kappa}(\tau,s_0)&=&\sum_i
f_{G_i}^2m_{G_i}^{2\kappa+4}e^{m_{G_i}^2\tau}, \ \ \kappa>-1.
\label{sr2}
\end{eqnarray}

Usually the lowest $\kappa$ value SRs are used to extract the value
of glueball mass because they are more stable with respect to the
continuum threshold
\begin{eqnarray}
m^2&=&\frac{S_1(\tau,s_0)}{S_0(\tau,s_0)}\nonumber\\
m^2&=&\frac{S_0(\tau,s_0)}{S_{-1}(\tau,s_0)+\Pi(0)}. \label{sr3}
\end{eqnarray}
In principle, one should obtain the same mass for the glueball in the above two SRs. It was
found that the extracted masses for the scalar glueball are quite different in the two.
However, as a recent analysis shows, if one includes the direct instanton contribution
the mass arising from the two SRs is the same.\cite{Forkel:2003mk,Forkel:2000fd,Harnett:2000fy}

These works also suggest a prominent role of the instantons in the binding of the $0^{++}$
glueball and show relations between the main properties of the $0^++$ glueball (mass and size), and
the bulk features of the instanton distribution.

In the meantime, the study of instanton contributions to SRs has been extended to other
glueball channels.\cite{Forkel:2003mk,Harnett:2000fy} While direct instanton contributions are
expected to be small in the tensor channel (mainly due to the absence of the leading
correction with respect to the small packing fraction of instantons in the QCD vacuum), they turn
out to be also important in the pseudoscalar channel. However, we should stress that the
instanton effects are counterbalanced in the pseudoscalar channel by a screening of the
topological charge.\cite{Forkel:2003mk} This screening occurs when the quark contribution is
important and leads to a prediction consistent with a low energy theorem. The most accurate
result , along these lines, for the masses of the scalar (S) and pseudoscalar (P) glueballs
including direct instanton effects in traditional SR calculations is
\begin{eqnarray}
m_S \approx 1.25 \text{ GeV,} \ \ \ m_P \approx 2.2 \text{ GeV}.
\nonumber
\end{eqnarray}
Within the uncertainties of the QCD sum rule approach these values
are in agreement with lattice calculation, which are
quenched calculations. Note that the SR predictions include
sea quark effect and thus one should be very careful when comparing
with the quenched lattice results. In the quenched approximation,
the topological charge screening disappears and the instanton effect
in the SR for the pseudoscalar glueballs is bigger.

We should emphasize  that one  central problem of glueball spectroscopy, namely the mixing
between quarkonium and the spin zero glueball states still has not been solved  so far. Recently,
some steps have been done in this direction,\cite{Kochelev:2005tu,Harnett:2008cw}  and
 it has been shown that the instantons induce
a strong mixing between the two states.\footnote{ The importance of the quark loop
contributions to the spin-zero gluonic correlators arising from instanton-antiinstanton
configurations has been demonstrated in refs. \cite{Schafer:1994fd,Tichy:2007fk}.} In this
respect we would like to point out that from our point of view, the tensor $2^{++}$ glueball
channel is the most clean gluonic channel. Indeed, for this state the leading instanton
contribution is zero, as follows from the structure of the interpolating current for this
channel, which is proportional to the energy-momentum tensor, Eq.~\eqref{tensor}, and from the
fact that the instanton is a vacuum gluon field with zero energy. Therefore, one can expect
very small tensor glueball mixing with quarkonium and, as a consequence, a small width for
this state and thus the possibility to separate it from the tensor quarkonium states. The SR
prediction for  the mass and decay constant of $2^{++}$ state is
\cite{Narison:1996fm,Narison:1999hg}
\begin{equation*}
m_T \approx 2.00 \text{ GeV},\quad \Gamma_{T \rightarrow \pi\pi+KK+\eta\eta}< (119\pm 36) \text{ MeV}.
\end{equation*}

Due to absence of  big uncertainties in the direct
instanton contribution and the small expected mixing with quarkonium
states in this channel, this SR prediction is
on more solid ground than the those for the zero spin glueball
states.

\section{Glueball Production and Decays}\label{sec:prodecay}

One of the most striking features of QCD is the prediction that glueballs might exist, a
prediction which has proven difficult to verify. We have seen how different theoretical
approaches describe their properties, and now we are going to discuss possible scenarios for
their experimental production and detection.

The strategy relies on a very few assumptions. Glueballs are extra
states, beyond the $\bar{q}q$ spectrum. To exploit this we must
understand the ``ordinary'' $\bar{q}q$ spectrum very well, using
data from J/$\psi$, $B$, and $Z$ decays, and from $\bar{p}p$, $\pi
p$, $\gamma \gamma$, and $\gamma N$ scattering. Glueballs are flavor
singlets so their decays should be $SU(3)_F$ symmetric, if the effects of violation
of $SU(3)_F$ symmetry are small. Since hybrid
and $q\bar{q}\bar{q}q$ states are also ``extra'' states and some are
flavor singlets, the only distinguishing property unique to
glueballs is their strong coupling to the color singlet digluon
channel.

A convincing way to see a glueball would be to detect a resonance
 whose quantum numbers are
not possible for mesons composed of quark and antiquark, {\it e.g.} $J^{PC} =
0^{--},0^{+-},1^{-+},2^{+-}$. These glueballs are sometimes called oddballs.
 They are
predicted to be rather narrow and easy to identify experimentally.\cite{Page:2001gs} These
states seem to appear for very large masses. The lightest oddball with
 $J^{PC} = 2^{+-}$ and
a predicted mass of $4.3$ GeV will be within range of the
 future experimental program at
$\overline{\mbox{P}}$ANDA.\cite{PANDA} It is conceivable that comparing
oddball properties
with those of non-exotic glueballs will reveal deep insight into the glueball structure since
their spin structures are different. However, if oddballs happen to appear at lower masses we
cannot rule out a glueball interpretation but they should be strongly mixed with hybrid
mesons.

\begin{figure}[ht]
\begin{tabular}{ccc}
\includegraphics[width=0.29\textwidth,height=32mm]{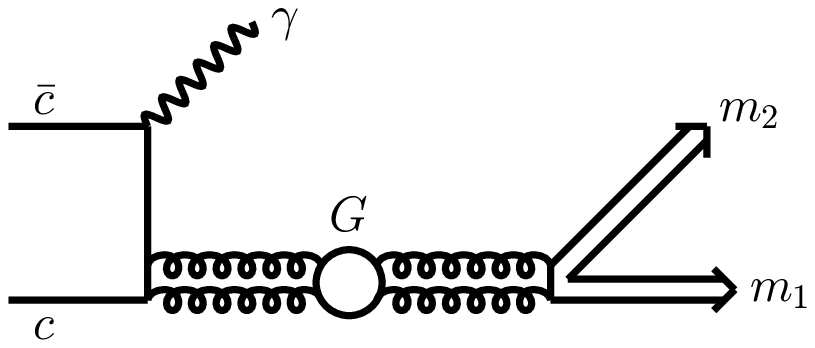}&
\hspace{3mm}\includegraphics[width=0.26\textwidth,height=29mm]{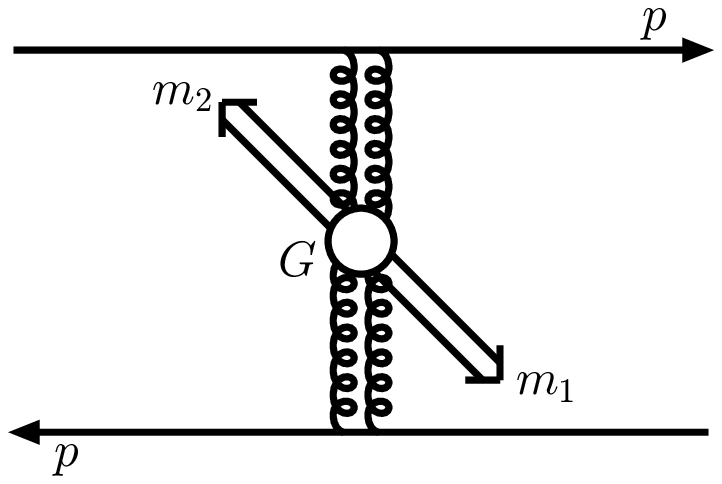}&
\hspace{3mm}\includegraphics[width=0.32\textwidth,height=31mm]{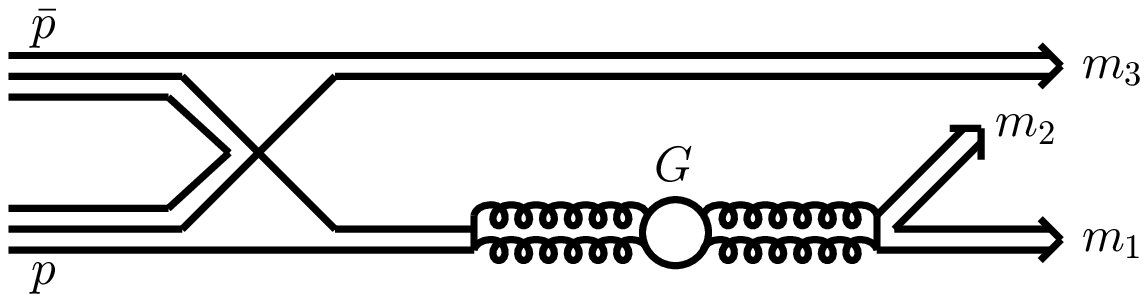}
\end{tabular}
\vspace*{-3mm} \caption{ Diagrams possibly leading to the formation
of glueballs: radiative J/$\psi$ decays, pomeron-pomeron collisions
in hadron-hadron central production, and in $ p\bar p$
annihilation.\label{fig:gluemod}}
\end{figure}

However, even non exotic glueballs can be identified by measuring an overpopulation of the
experimental meson spectrum and by comparing masses, quantum numbers and decay channels, with
predictions from models or lattice QCD. The best scenario is to look for Zweig forbidden
processes, since the decay into glueballs is dominant in these cases.\cite{Robson:1977pm}
These processes are known under the name of gluon-rich processes, some are depicted in
Fig.~\ref{fig:gluemod}. Let us specify some of the processes that fulfil these requirements
and are optimal experimental scenarios to search for glueballs:

\begin{itemize}

\item [(i)] {\bf J/$\psi$ decays:}
the most suggestive process is the radiative J/$\psi$ decay. The J/$\psi$ is
narrow; the
$D\bar D$ threshold is above the mass of the J/$\psi$ and the OZI rule
 suppresses decays of
the $c\bar c$ system into light quarks. In most decays, the J/$\psi$ undergoes
 a transition
into 3 gluons which then convert into hadrons. This is a high multiplicity process difficult
to detect. But J/$\psi$ can also decay into 2 gluons and a photon. The photon can be detected,
the two gluons interact and must form glueballs. BES III will provide
 huge J/$\psi$ data
samples allowing definitive studies of J/$\psi$ decay and, especially,
 partial wave analysis
of the glueball-preferred radiative J/$\psi$ decay channel.\cite{BES}

\item [(ii)]{\bf Central production:} glueballs decay into hadrons and hence
hadro-production of glueballs is always possible. Central production
is a process in which glueballs should be abundantly produced.\cite{Kochelev:1999kk} In
central production two hadrons scatter diffractively, gluon-gluon
fusion processes should be abundant and no valence quarks are
exchanged. The process is often called double pomeron exchange, with
the pomeron presumed to be a multi-gluon color singlet. The absence
of valence quarks in the production process makes central production
a good place to search for glueballs.

\item [(iii)] {\bf $\bar pp$  annihilation:} quark-antiquark pairs annihilate into
gluons, they interact and may form glueballs, {\it e.g.} $\bar pp \rightarrow \pi^0 G$. A new
era of precise experiments will start when $\overline{\mbox{P}}$ANDA starts to
operate.\cite{PANDA}. The energy range of this experimental program will allow the study of
oddballs. Glueballs can either be formed directly in the {\bf $\bar pp$}-annihilation process,
or produced together with another particle. In both cases the glueball decay into final states
like $\phi \phi$ or $\phi \eta$ would be a favorable reaction below $3.6$ GeV, while J/$\psi
\eta$ and J/$\psi \phi$ are the first choice for more massive states.

\item[(iv)] {\bf $\gamma \gamma$ fusion:} production of glueballs should be suppressed in
this case since photons couple to the intrinsic charges. So we
should expect a glueball to be strongly produced in radiative
J/$\psi$ decays but not in  $\gamma\gamma$ fusion. Radial
excitations might be visible only weakly in J/$\psi$ decays but they
should couple to $\gamma\gamma$. However, if a very low mass
glueball would exist, this would be one of the cleanest ways to find
it in a difficult experiment due to the low counting rate.

\item[(v)] {\bf photoproduction: } it has been a continuous source of
meson resonances. Photoproduction is expected to be particularly effective in producing exotic
hybrids and glueballs. The photon, {\it via} Pomeron exchange from the vacuum, or {\it via} its hadronic
component, can create states with exotic $J^{PC}$. These mesons can be hybrids, glueballs or
mixed states. Moreover, also non exotic hybrid (glueball) states can be produced which
manifest themselves as extraneous states that cannot be accommodated within $\bar q q$ nonets.
However, there is little data on photoproduction of light mesons. This will drastically change
when the GlueX experiment goes into operation, since it is designed to collect data of
unprecedent statistics and quality.\cite{GlueX} Moreover,  the (linear) polarization of the
beam will allow the identification of (exotic) quantum numbers and the determination of the
details of the production mechanism of (glue) mesons.
\end{itemize}\vskip 0.3cm

\vskip 0.3cm

The status of glueball observation would become clearer with a combination of data of $ p p$
and $e^+ e^-$ machines with large statistics, however since the glueball width into
two-photons is small, this would require very high luminosity. It has been pointed out that
heavy-ion colliders, due to their large center of mass energy, allow access to the Regge
region $s\gg|t|$ and hence to the production of glueballs in peripheral collisions through
photon-photon collisions as well as double pomeron exchange
\cite{Natale:1995ap,Schramm:1999tt,Bashkanov:2002qu}.

Distinctive features can be derived from the decays of the glueballs
since they are flavour singlets: decays to $\eta\eta^{\prime}$
identify a flavour octet and radiative decays of glueballs are
forbidden. However, these arguments have to be taken with care since
mixing of a glueball with mesons, having the same quantum numbers,
can occur and would dilute any selection rule.

An interesting idea recently proposed is that of chiral suppression.\cite{Chanowitz:2005du} If
chiral symmetry breaking in glueball decay is dominated by quark masses, then the coupling of
a spin zero glueball to light $q\bar{q}$ pairs is chirally suppressed. An interesting
consequence is that radiative J/$\psi$ decay becomes a filter for new physics in the $J=0$
channel, since at leading order radiative decays to spin zero light quark mesons are
suppressed while radiative decays to $J=0$ glueballs, hybrid, and four quark states are not
suppressed. This idea is not without controversy. \cite{Chao:2007sk,Chanowitz:2007ma}

The glueball groundstate $J^{PC} = 0^{++}$ would be an ideal state to find since it has many
fundamental
connotations.\cite{Novikov:1979va,Migdal:1982jp,Ellis:1984jv,West:1996du,Vento:2004xx} In
gluodynamics, as we have seen, the situation that arises from lattice calculations is clear
and the masses of the scalar glueballs are large $m > 1$~GeV. However, when sea quarks are
considered no firm conclusion about the scalar spectrum can be drawn. The theoretical
calculations based on QCD sum rules and/or low energy theorems lead to contradictory results.
Its properties, {\it i.e.}, mass, decay channels and widths still differ among the various
approaches.

Experimentally there are too many isoscalar, scalar mesons between 0.6 and 1.75 GeV to be
explained by the naive quark model alone.  From the theoretical point of view there has been a lot of
 debate about the structure of these states. Let us summarize here the different
 theoretical explanations for the spectrum.
 Jaffe obtains  the $f_0(600)$ and the $f_0(980)$ in
the bag model as cryptoexotic $q\bar{q}\bar{q}q$ states.\cite{Jaffe:1976ig}
The p-wave $q\bar{q}$ scalar
nonet is likely to lie in the region of the other spin-triplet p-wave nonets, with isoscalars roughly
 between $\sim 1250$ and $\sim 1600$~MeV. \cite{Amsler:1995tu}
Dominguez
and Paver,\cite{Dominguez:1986td} Bordes, Pe\~narrocha and Gim\'enez,\cite{Bordes:1989kc} and
Kisslinger and Johnson,\cite{Kisslinger:2001pk} obtain, using low energy theorems and/or sum
rule calculations,  a low lying and narrow glueball
(mass $<$ 700~MeV and $\Gamma_{\pi\pi} <$ 100~MeV). Vento using low
energy theorems has also proposed the existence of a low mass ($ m < 700$~MeV) and
almost sterile scalar glueball.\cite{Vento:2004xx}
Narison and collaborators,\cite{Narison:2002gv,Dosch:2002hc} using sum
rules, get a broader (200-800~MeV), heavier (700-1000~MeV) scalar glueball in this range.
Forkel,\cite{Forkel:2003mk}
obtains a scalar glueball at 1250~$\pm$~200~MeV with a large width ($\sim$300~MeV).
Between $\sim
1400$ and 1750~MeV there are three $I,J^{PC}= 0,0^{++}$ states: $f_0(1370)$, $f_0(1500)$, and
$f_0(1710)$.
In the analysis of Vento \cite{Vento:2004xx}  two scalar glueballs appear in this range, an
intermediate mass glueball ($\sim 1300$ MeV)  and another one in the upper range
($\sim 1500-1700$ MeV).
The narrow state at 1500~MeV, discovered in $p
\bar{p}$ annihilation by Crystal Barrel (CB) is considered by Amsler and Close a good
candidate for the glueball groundstate\cite{Amsler:1995tu,Amsler:1995td,Abele:2001js} while
Chanowitz considers the $f_0(1710)$ as the scalar
groundstate\cite{Chanowitz:2006wf,Chanowitz:2005du}. Finally, that the scalar glueball is
shared by both resonances is also contemplated.\cite{Chanowitz:2006wf}

\section{Glueballs and the Quark-Gluon Plasma}\label{sec:qgp}

Relativistic heavy-ion collisions might be a tool to produce
glueballs. It is conceivable that glueball production becomes a
dominant part in central nucleon collisions.

Two scenarios will be analyzed:

\begin{itemize}

\item [(i)]{\bf Quark Gluon Plasma phase}: one expects that at some point
above a certain critical temperature a plasma of quarks and gluons, named quark-gluon plasma
(QGP) is formed.\cite{Shuryak:1980tp} This phase is characterized by a large amount of thermal
gluons. As the QGP cools down the gluons can form singlet configurations via the color
interaction.

\item [(ii)] {\bf Strong Coulomb phase}: a
recent formulation of the dynamics in the region above the transition temperature $T_C$, based
on a description of recent experiments in ultra relativistic heavy ion collisions,\cite{RHIC}
states that, despite de-confinement, the color Coulomb interaction between the constituents is
strong and a large number of binary (even color) bound states, with a specific mass pattern,
are formed.\cite{Shuryak:2004tx} This phase we call Strong Coulomb Phase (SCP). The QGP phase
occurs at a much higher temperature $T_{QGP} > (2-3) T_C$ when the bound states dissolve.

\end{itemize}

Let us describe some attempts to find signatures arising from QGP. The basic idea is that
during the hadronization process in this gluon rich environment the gluons combine due to the
strong force into glueballs. As the QGP cools further and transforms into hadronic matter,
these glueballs decays into conventional hadrons giving rise to signatures of their existence.
It has been claimed that the existence of glueballs alters the $K/\pi$ ratio in the final
state.\cite{Gao:1995ni}

It is conceivable that glueball production becomes dominant in central nuclear collisions
since the existence of QGP provides a gluon rich environment especially at high energy
density. In this scenario the lowest mass glueballs are copiously produced. Particular decay
modes $0^{++}, 2^{++} \rightarrow K \bar{K}$ and $0^{++}\rightarrow \pi^+ \pi^- l^+l^-$ have
been investigated\cite{Kabana:1999jn} and enhancements associated with possible glueball
production observed. Search strategies, including dipion production have been also
proposed.\cite{Minkowski:2000fu}

Estimates within thermal models for the multiplicities of scalar glueballs in central Au-Au
collisions at present and future experimentally available energies, {\it i.e.} from AGS to
LHC, have been presented.\cite{Mishustin:2006wm} For the experimental identification of
glueballs one can use the decay modes $G\rightarrow K \bar{K}$, $G\rightarrow \gamma \gamma$
and $G\rightarrow 2\pi l^+l^-$. Despite of small branching ratio, the $2\gamma$ channel has
the important advantage that photons have practically no rescattering in the hadronic medium.
This analysis leads to maximal multiplicities for the glueball of $ 1.5-5\%$ of the $\phi$
meson multiplicity. Even larger yields are expected in the case of explosive hadronization of
the quark gluon plasma.\cite{Mishustin:2006wm}

Let us now turn to the second scenario. From its inception two decades ago, the high-T phase
of QCD commonly known as the Quark-Gluon Plasma,\cite{Shuryak:1980tp} was described as a
weakly interacting gas of ``quasiparticles''(quarks and gluons). Indeed, at very high
temperatures asymptotic freedom causes the electric coupling to be small and the QGP to be
weakly interacting. At intermediate temperatures of few times the critical temperature $T_c$
of immediate relevance to current experiments, there is new and growing evidence that the QGP
is not weakly coupled. In this region QCD seems to be  close to {\em a strongly coupled
Coulomb regime}, with an effective coupling constant $\alpha\approx$ 0.5-1 and multiple {\em
bound states} of quasiparticles.\cite{Shuryak:2004tx} This phase we call Strong Coulomb Phase.
This description is not universally accepted since some lattice calculations do not to find
bound states above the transition.\cite{Koch:2005vg,Karsch:2005ps} Also, one must be aware
that other explanations of the data have been presented.\cite{Ratti:2006wg} In order to
clarify these issues it is interesting to find clear experimental observables that would shed
some light into the discussion.

\begin{figure}[htb]
\begin{minipage}[t]{6.0cm}
\hskip-0.0cm \epsfig{file=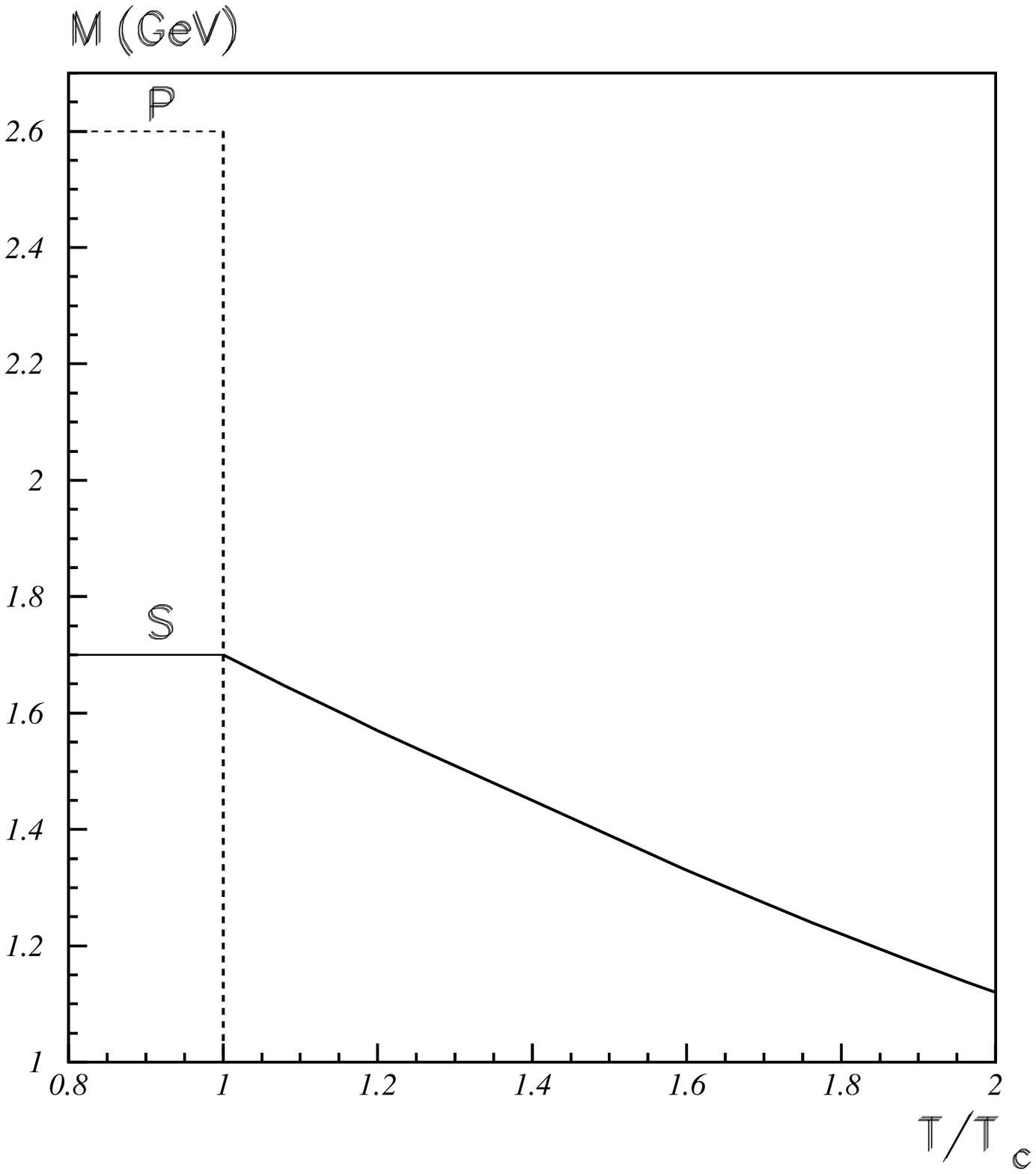,width=6.0cm,angle=0}
\end{minipage} \hskip 0.5cm \begin{minipage}[t]{6.0cm}
\epsfig{file=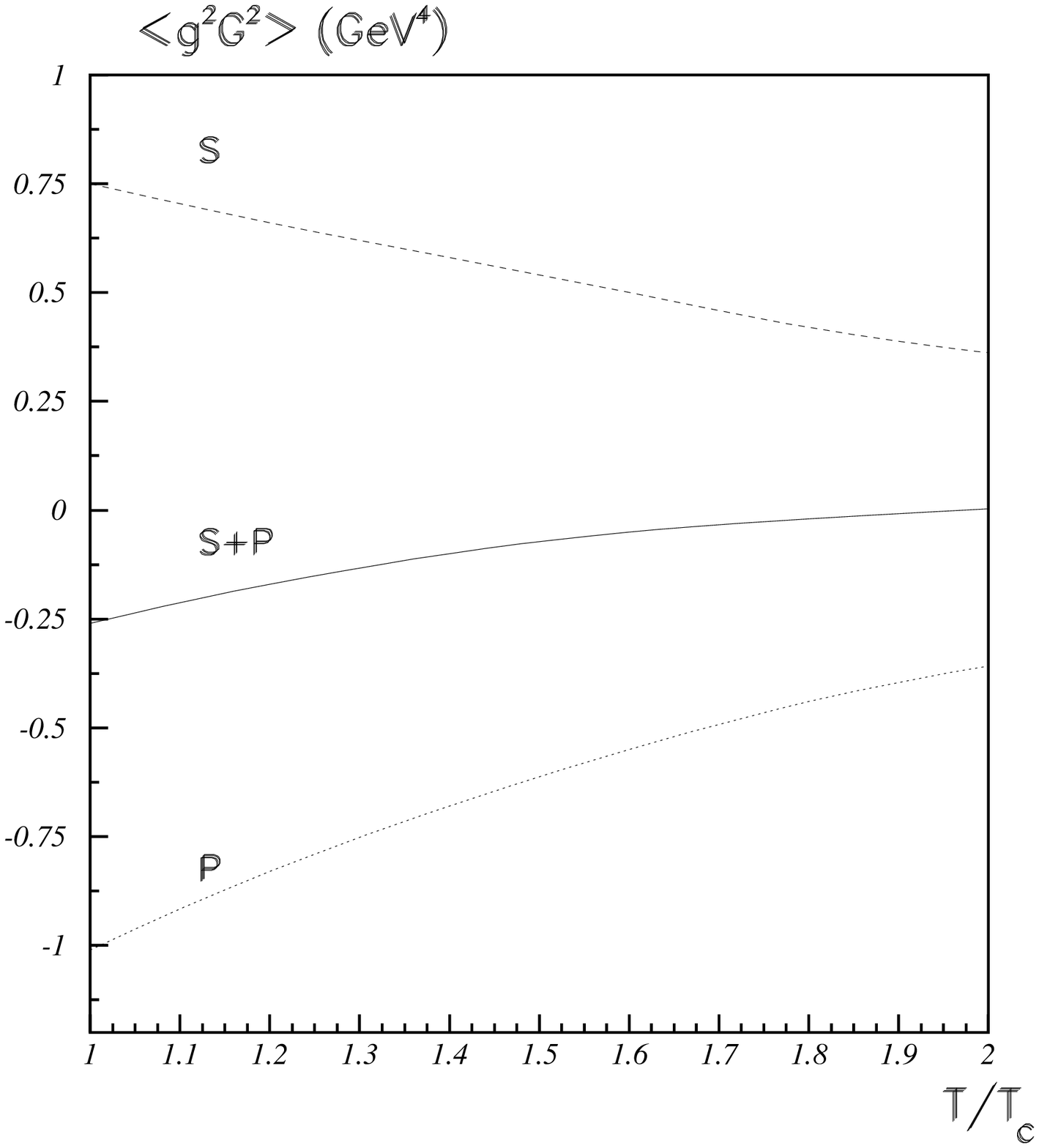,width=6.0cm,angle=0}
\end{minipage}
\caption{The solid (dotted) line is the  temperature dependence of
scalar (pseudoscalar) glueball mass (left).The temperature
dependence of the gluon condensate at $T>T_c$. The solid line is
total condensate, the dashed (dotted) line is the scalar
(pseudoscalar) glueball contributions (right).\label{glueball}}
\vskip 0.5cm
\end{figure}

Kochelev and Min consider the properties of scalar and pseudoscalar glueballs in
SCP.\cite{Kochelev:2006sx} In an effective Lagrangian approach, based on the low-energy QCD
theorems, they  find out that scalar glueball remains massive above deconfinement temperature
(see Fig. \ref{glueball}). At the same time, pseudoscalar glueball changes its properties in
QGP in a drastic way. Indeed, this glueball becomes massless at $T>T_c$ and therefore it can
contribute strongly to,the bulk properties of the plasma (see Fig. \ref{glueball}). They
demonstrate that the disappearance of pseudoscalar glueball mass above the deconfinement
temperature and its strong coupling to gluons gives the rise to the sign change of the gluon
condensate in the pure $SU(3)_c$ gauge theory as observed in the lattice calculations at
$T\approx T_c$ (see Fig. \ref{glueball}).\cite{Miller:2006hr} The strong nonperturbative
coupling of the glueball to the gluons leads to the conjecture that one might expect that the
role of very light pseudoscalar glueball in QGP must be quite similar to the role played by
the massless pion in nuclear matter below deconfinement temperature.

In SCP despite de-confinement the color Coulomb interaction between the constituents is strong
and a large number of binary (even color) bound states, with a specific mass pattern, are
formed.\cite{Shuryak:2004tx} With this input, the scenario envisaged by Vento for
gluodynamics, {\it i.e.} the theory with only gluons goes as follows.\cite{Vento:2006wh} The
strong Coulomb phase is crowded with gluon bound states and the lightest is the scalar
glueball, labelled $g$. As one moves towards the dilution limit, the binding energy of these
states decreases, the gluon mass increases, and therefore the color and singlet bound states
increase their mass softly until the gluons are liberated forming a
liquid.\cite{Shuryak:2004tx} However, as the system cools towards the confining phase, color
and singlet states decay into the conventional low lying glueballs, in particular $g$. Thus
the number of $g$'s becomes large.

This reasoning generalizes to QCD since in the SCP the multiplicity of glueball channels is
larger than in the confined phase. The ratio of glueball to meson channels goes from 1 to 8
below the phase transition to 1 to 2 above.\cite{Shuryak:2004tx} Thus  the number of scalar
glueballs is much larger in SCP than in the cold world.

As the fireball cools a ``large number" of gluonic bound states
decay by gluon emission into $g$'s. The emitted gluons form new
bound states of lower mass due to the strong color Coulomb
interaction. As we approach the confinement region the mass of the
color bound states increases and it pays off to make multiparticle
color singlet states, which decay by rearrangement into ordinary
color singlet states. Since the coupling is strong and the phase
space is large, these processes take place rapidly. Thus in no time,
close to the phase transition temperature $T_C$, a large number of
scalar glueballs populate the hadronic liquid.

\vskip 0.5cm
\begin{figure}[htb]
\centerline{\epsfig{file=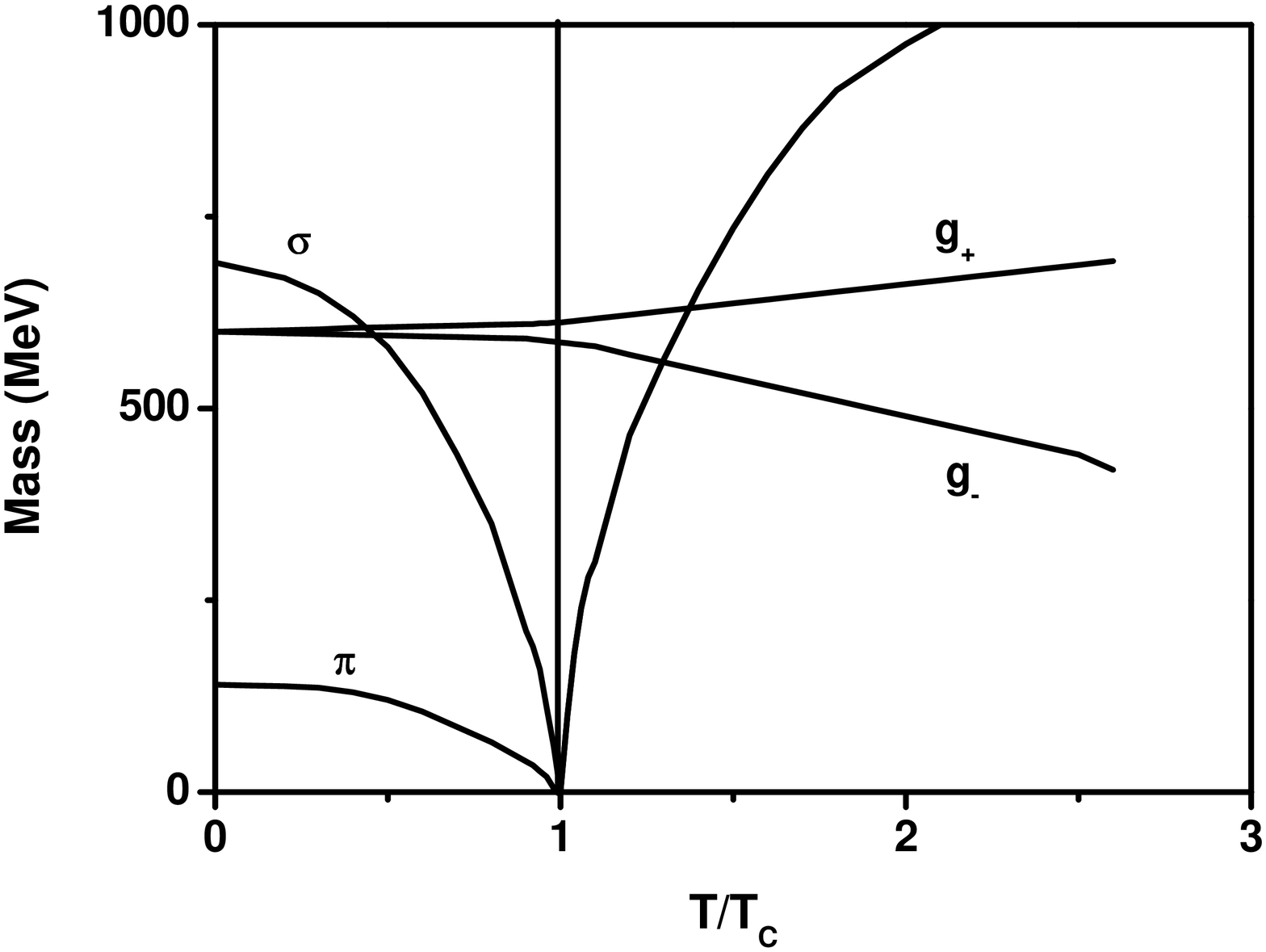,width=6.5cm,angle=270}} \caption{Behavior of the
masses of $\sigma$,$\pi$ and $g$ across the QGP phase transition according to model
calculations. $g_\pm$ label the upper(lower) limits of the $g$ in mass model
calculations.$^{130}$ \label{SCP}} \vskip 0.5cm
\end{figure}

Let us now analyze the experimental signal of these phenomenon. If we assume that the $\sigma$
is the O(4) partner of the $\pi$ in the chiral symmetry realization of QCD, its mass decreases
when approaching the phase transition, becoming degenerate with the pion at $T_C$ (see
Fig.~\ref{SCP}). Beyond $T_C$, in the SCP, chiral symmetry is restored, and $\pi$ and $\sigma$
remain degenerate for $T > T_C$. Thus in the SCP the $\sigma$ can only decay in $2\gamma$ for
obvious kinematical reasons. The glueball $g$ does not vary its mass in this region
appreciably. Thus even before we reach $T_C$, the mixing between $g$ and $\sigma$ disappears
(see Fig.~\ref{SCP}) and $g$ becomes stable around $T_C$. However, in the SCP the mass of the
$\sigma$ increases and in a certain region of $T$ it again becomes degenerate with $g$ and
mixing is restored. Thus the physical  $g$ is able to decay, once the $\sigma$ component is
attained, only to $2\gamma$.

\begin{center}
\begin{figure}[htb]
\centerline{\epsfig{file=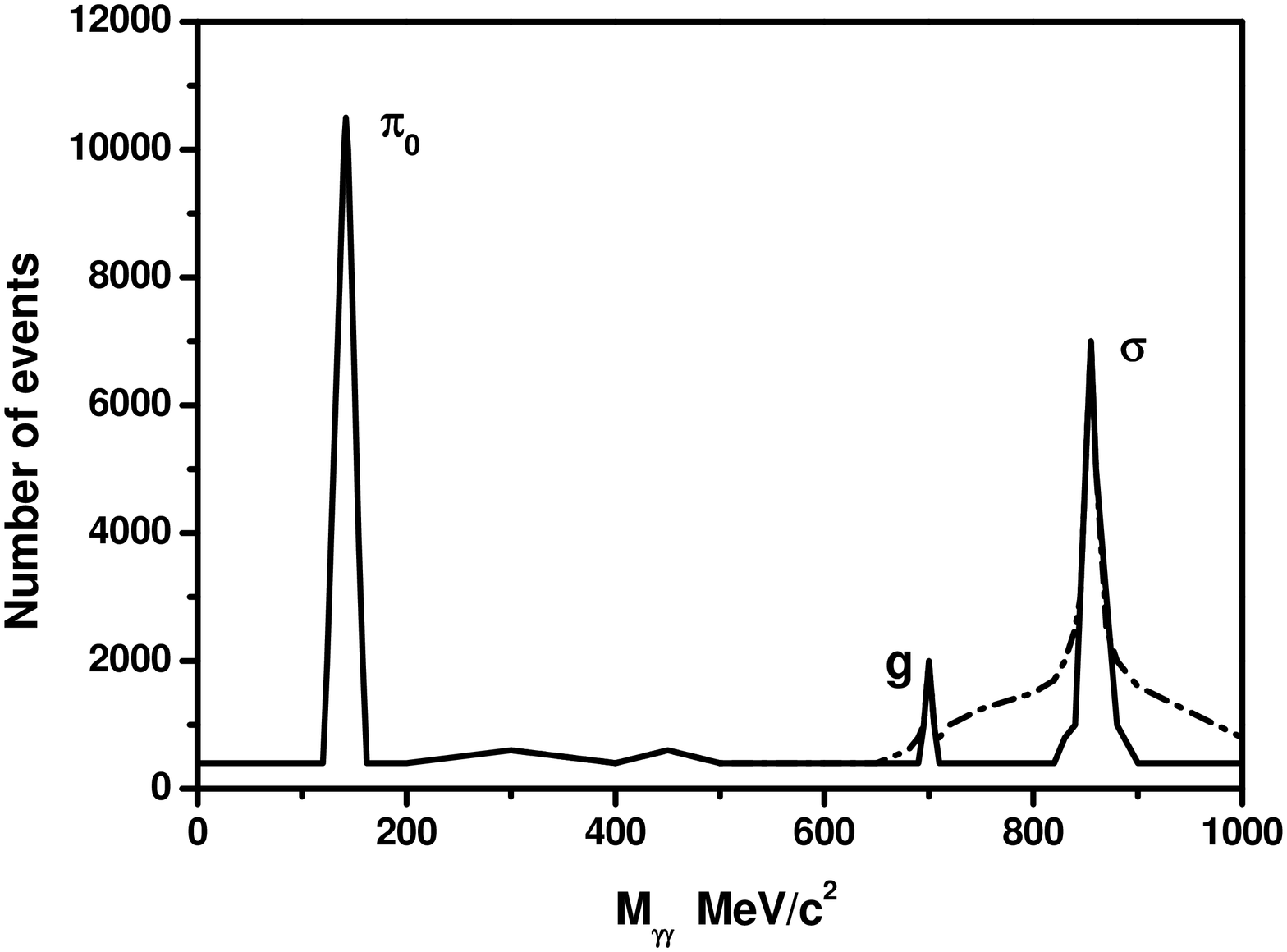,width=.50\linewidth,angle=270}} \caption{Expected fit to
the two-photon invariant mass spectrum in central Pb-Pb collisions after substraction from the
background. The $2\gamma$ decays should allow for a clear separation of $g$ and $\sigma$. The
figure includes an estimate of the effect of the hadronic widths on the fits (dot-dashed
line). \label{fits}} \vskip 0.5cm
\end{figure}
\end{center}

The enhancement in the number of $g$s with respect to the hadronic phase arises because of the larger
population of glueballs in the SCP, as described above, and because these particles are stable in the medium
against the dominating hadronic decays (see Fig.~\ref{SCP}). Thus a clear signal for the
existence of a low mass scalar glueball and a confirmation of the SCP scenario would be two
$2\gamma$ peaks corresponding to $g$ and the $\sigma$-meson as shown in Fig.~\ref{fits}.

The investigation on jets in the relativistic heavy ion collisions at RHIC provides a deep
insight into the properties of the quark-gluon plasma.\cite{RHIC} One of the important RHIC
discoveries is the jet quenching phenomenon coming from the partonic energy loss in QGP. In
the conventional approach to the jet quenching the perturbative (pQCD) type of energy loss is
taken into account by the channels, elastic and radiative, of one-gluon exchange between the
jet and the massless gluons and quarks.\cite{Wicks:2007zz}
However, the large quark-gluon rapidity
 density $dN_{qg}/dy\approx 2000$ which is needed to describe the RHIC jet quenching
 data within this approach, seems to be in contradiction with the
 restriction $dN_{qg}/dy\leq 1/4dS/dy\approx 1300$ coming from the
measured final entropy density $dS/dy\approx 5000$.\cite{Muller:2006ee} Furthermore, the
lattice calculations show that even at very high temperature gluons and quarks still interact
strongly in QGP. \cite{Miller:2006hr}

\begin{figure}[htb]
\begin{minipage}[t]{6.0cm}
\hskip-0.0cm \epsfig{file=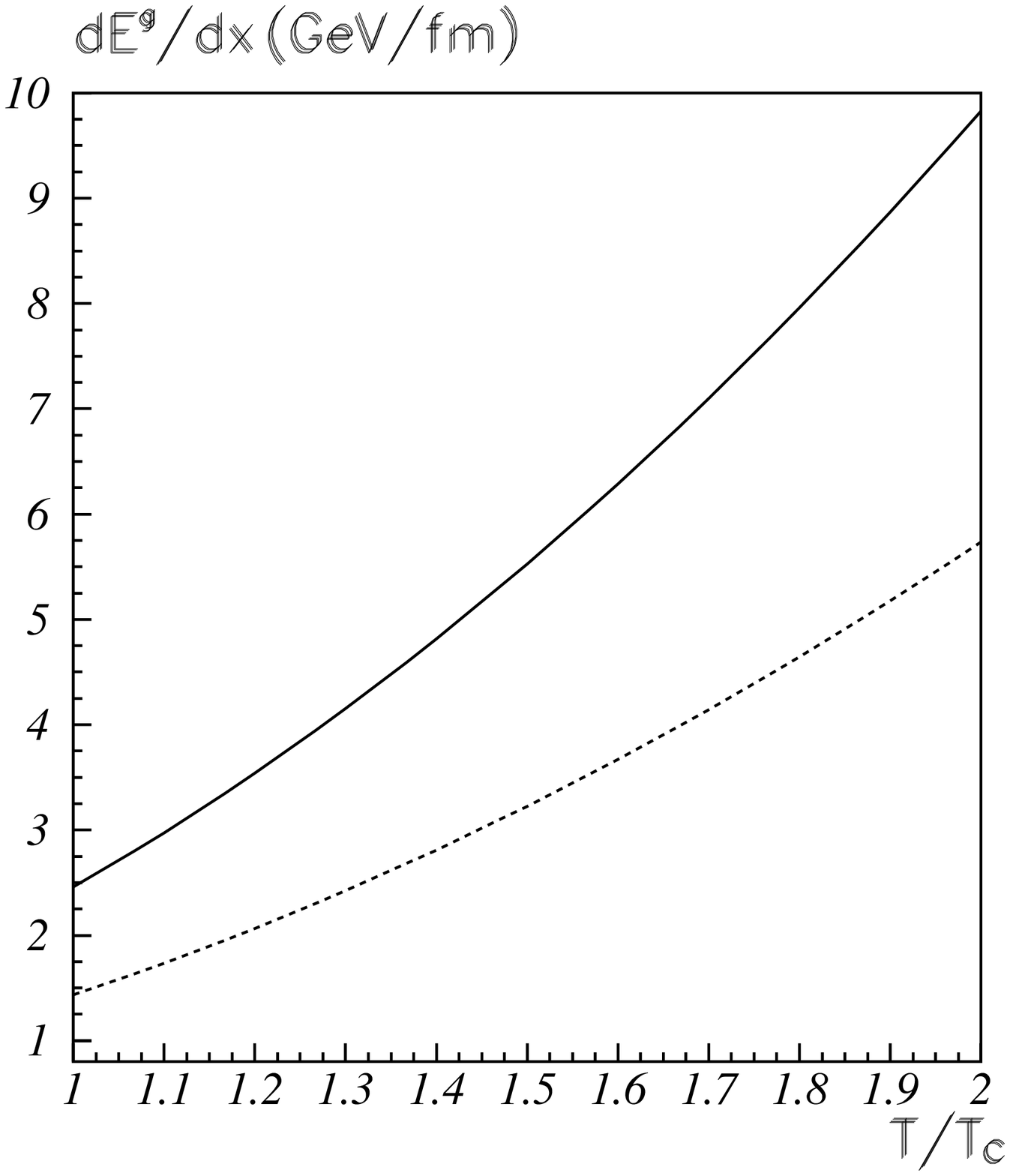,width=6.0cm,angle=0}
\end{minipage} \hskip 0.5cm \begin{minipage}[t]{6.0cm}
\epsfig{file=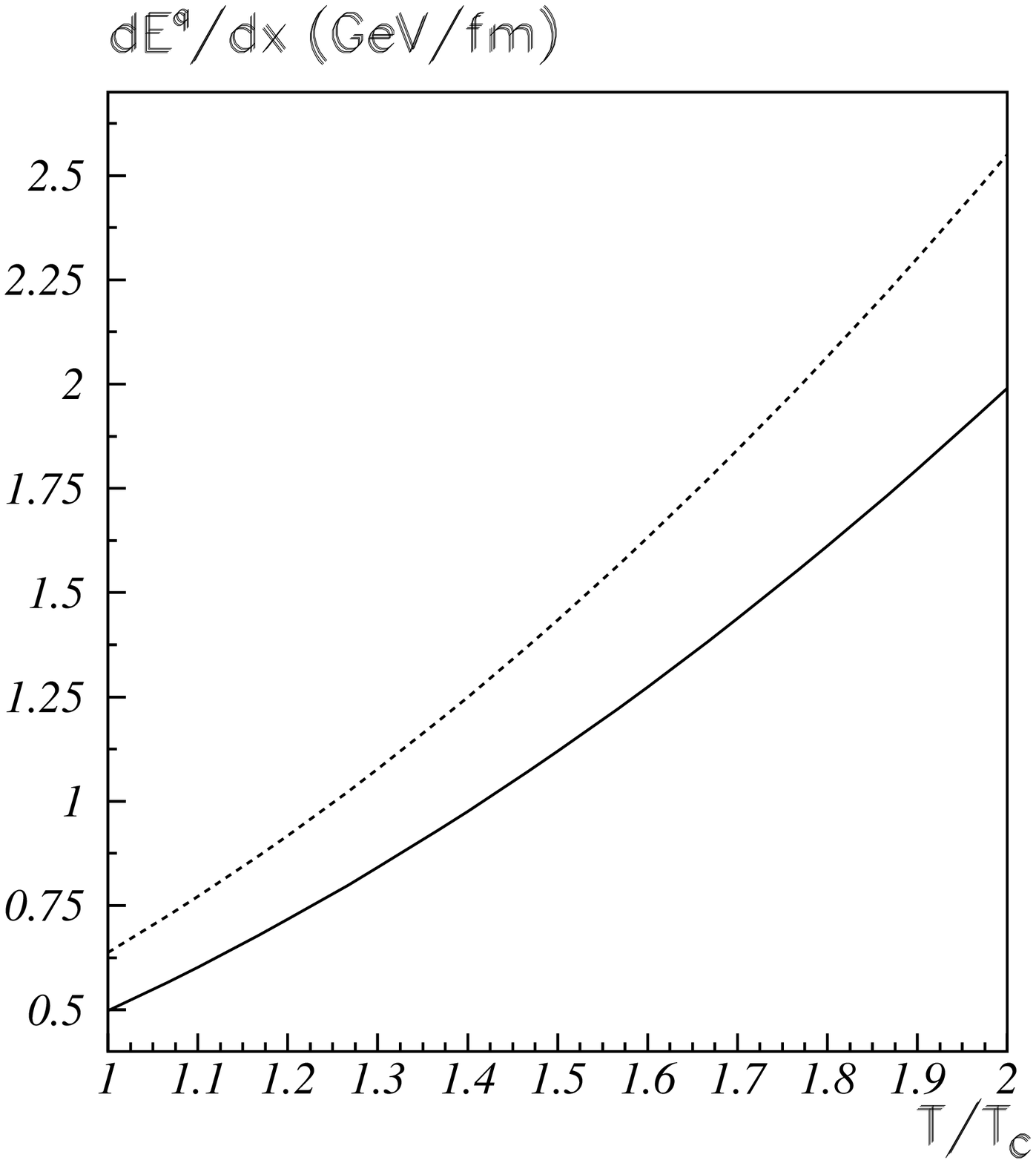,width=6.0cm,angle=0}
\end{minipage}
\caption{The temperature dependence of gluon energy loss (left). The
temperature dependence of quark energy loss (right).The solid
(dashed) line is glueball (pQCD) contribution .\label{t1t2}}
\end{figure}

As seen above, it has been  suggested that the glueballs, the bound states of gluons, can
exist above deconfinement temperature and may play an important role in the dynamic of
strongly interacting QGP.\cite{Kochelev:2006sx,Vento:2006wh} In particular,
it has been suggested that a very light pseudoscalar glueball can exist in
QGP and might be responsible for the residual strong interaction between gluons.\cite{Kochelev:2006sx}  The lattice
results showing a change of sign of the gluon condensate\cite{Miller:2006hr} and a small value
of the topological susceptibility above $T_c$ can be explained in the glueball picture as
well. Furthermore, one expects that the suppression of the mixing between glueballs and
quarkonium states in the QGP leads to a smaller width for the former as compared to the
vacuum.\cite{Vento:2006wh} This property opens the possibility for a clear separation of the
glueball and the quark states in heavy ion collisions. Such separation is rather difficult in
other hadron reactions  due to existence of strong glueball-quarkonium mixing in the vacuum.

Min and Kochelev  made an estimate of the energy loss induced by interaction of an energetic
parton,  which was produced in the hard scattering of two heavy ion's partons, with glueballs
in the hot quark-gluon plasma.\cite{Alles:1996nm} They showed that such contribution leads to
a significant energy loss as can be seen from Fig.~\ref{t1t2}. In particular, for the gluon
jet such contribution is about a few GeV/fm and approximately twice larger than the
perturbative elastic loss.\cite{Peshier:2006hi} It should be pointed out that more than one
half of contribution to the gluon energy loss comes from interaction of gluon with the light
pseudoscalar glueball in QGP. Thus in spite of the fact that for the quark jet the glueball
contribution is smaller than perturbative elastic loss, it can not be neglected in comparison
with latter one. The existence of such light bound state of gluons above $T_C$ is crucial for
the understanding of the large observed partonic energy loss in QGP. Therefore, not only pQCD
type of energy loss but also glueball-induced loss, arising from existence of scalar and
pseudoscalar glueballs in QGP, are important for the understanding of the RHIC results such as
the jet quenching.

\section{Other Developments}\label{sec:other}

\subsection{Glueballs and the Pomeron }\label{sec:pomeron}

It is well known that in the many high energy reactions  with small
 momentum transfer
the exchange by the
highest-lying Regge trajectory, called as a soft Pomeron, gives the
 dominant contribution. This exchange carries vacuum quantum numbers and has
very peculiar properties in comparison with the usual Regge pole trajectories as the
 $\rho$, $\pi$, {\it etc}. From the
analysis of the varions cross sections it follows that the
Pomeron trajectory has a linear behavior
\begin{equation}\label{eq:pomeron}
    J(t=m^2) = 1.08 + 0.25 m^2.
\end{equation}
The Pomeron does not seem to be related to the usual mesons since the
latter have  usually lower intercepts and
very different slopes. There has been a long-standing
speculation that the physical particles on the trajectory might be
glueballs.

Meyer and Teper investigated the pure gauge spectrum on a lattice to
check its compatibility with a linear trajectory.\cite{Meyer:2004jc}
They calculated the lightest states $J^{PC}=2^{++}, 4^{++}, 6^{++}$
since the trajectory has an even signature. Their masses (within
errors bars) are consistent with the Pomeron
trajectory~\eqref{eq:pomeron}. These studies of quarkless QCD helps to
understand the close relation between the Pomeron and the pure glue
states, but it is worth mentioning that the physical particles
should probably mix with quarkonia. The linearity of the trajectory,
supported by these arguments, favors flux tube and stringy pictures as
discussed in the constituent model section,  at least for (even)$^{++}$
glueballs.

\subsection{Glueball - $ q \bar{q}$ mixing}

Most of our review has been dedicated to the study of pure glueball
states. We have assumed in our discussion that glueballs are
physical objects, which is fine for the purpose of the developments
thus far. But in order to establish contact with reality we have to
discuss in some detail mixing, an idea which has been mentioned in
passing on several occasions. We first discuss the mathematics of
mixing in a two state  model and then apply the reasoning to the
physical reality in the isosinglet scalar sector.

\begin{figure}[htb]
\centerline{\epsfig{file=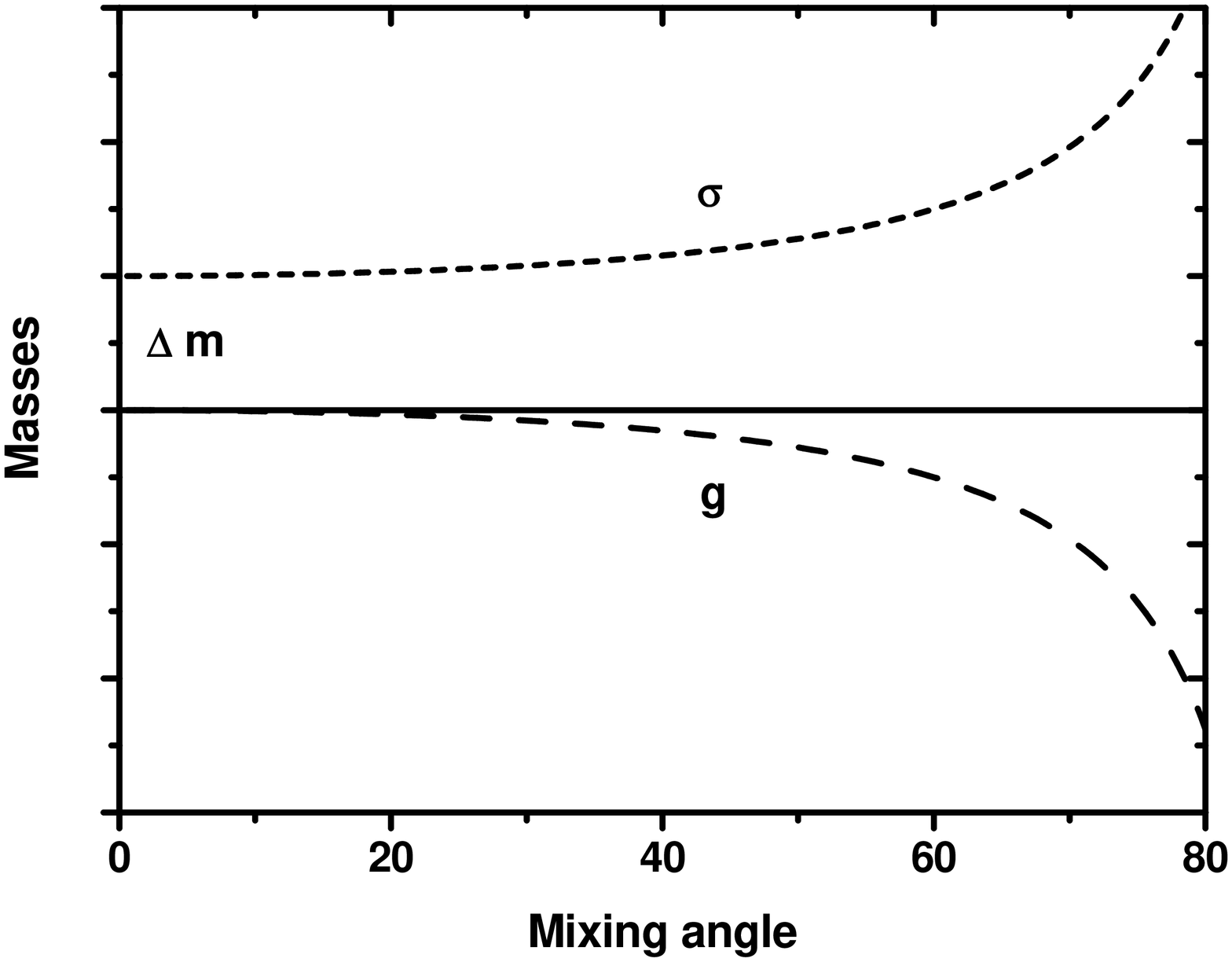,width=0.50\linewidth,angle=270}}
\caption{The limiting values for the masses of the physical $g$
(solid--dashed lines) and  $\sigma$ (solid--short-dashed lines) are
shown as a function of the mixing angle.}
\label{2statemixmass}\vskip 1cm
\end{figure}

Mixing arises when the total Hamiltonian is not diagonal in the glueball and $q \bar{q}$ Fock
space. In this case the physical states come out as superpositions of glueballs and $q
\bar{q}$ states. Let us assume that we have a pure glueball, we label $g$, and a $q \bar{q}$
state, we label $\sigma$, which have the same quantum numbers and, for simplicity, also the
same mass $m$, in a certain approximation. We next relax this approximation and the
Hamiltonian in the reduced $g$ and $\sigma$ Fock space becomes
\begin{equation} \left(
\begin{array}{c c}
m & \delta \\
\delta & m +\Delta m \\
\end{array} \right)
\end{equation}
The diagonal basis of this Hamiltonian can be presented as,
\begin{eqnarray} |\tilde{g}\rangle & =  & \cos{(\theta/2)}|g\rangle  - \sin{(\theta/2)}|\sigma\rangle
,
\\ |\tilde{\sigma}\rangle & =  &  \sin{(\theta/2)} |g\rangle + \cos{(\theta/2)} |\sigma\rangle ,
\end{eqnarray}
where the tilde labels the physical particles and $\theta$ is the
mixing angle.

The masses of the physical particles become
\begin{align} m_{\tilde{g}}& = m + \frac{\Delta m}{2} - r, &
m_{\tilde{\sigma}} & = m + \frac{\Delta m}{2} + r,  \label{masses}
\end{align}
where
\begin{equation} \tan{\theta} = \frac{2 \delta}{\Delta m}
\end{equation}
and
\begin{equation} r = \frac{\Delta m}{2} \sqrt{ 1 + \left(\frac{2\delta}{\Delta
m}\right)^2} = \frac{\Delta m}{2 \cos{\theta}}. \end{equation}

In Fig. \ref{2statemixmass} we represent the masses of the physical
states as a function of the mixing angle $\theta$. The curves
separate possible mass regions. The curves show that the two state
mixing scenario for positive $\Delta m$ leads to a ``light" glueball
and heavier meson. The opposite result can be obtained with a
negative $\Delta m$. The procedure can be generalized to more states
and to different initial masses as we next see. This mechanism has
been used to try to disentangle the complex isoscalar sector.

Vento has shown that a low mass, almost invisible isosinglet scalar glueball, could hide in
the low energy spectrum due to small mixing with the $f_0(600)$, and pointed out the possible
existence of other glueballs components among the remaining isosinglet scalars $f_0(1370)$,
$f_0(1500)$ and $f_0(1710)$.\cite{Vento:2004xx}

Close {\it et al.} have discussed the content of the the isosinglet scalar mesons $f_0(1370)$,
$f_0(1500)$  and $f_0(1710)$ in an attempt to discover which of them is dominantly a scalar
glueball.\cite{close1,close2} These authors have suggested that $f_0(1500)$ is primarily a
scalar glueball, due partly to the fact that $f_0(1500)$, discovered in $p\bar{p}$
annihilation at LEAR, has decays to $\eta\eta$ and $\eta\eta'$ which are relatively large
compared to that of $\pi\pi$\cite{ams95} and that the earlier quenched lattice
calculations\cite{Bali,Teper} predict the scalar glueball mass to be around $1550$ MeV.
Furthermore, because of the small production of $\pi\pi$ in $f_0(1710)$ decay compared to that
of $K\bar K$, they claim that the $f_0(1710)$ is primarily $s\bar s$ dominated. In contrast,
the smaller production rate of $K\bar K$ relative to $\pi\pi$ in $f_0(1370)$ decay leads to
the conjecture that $f_0(1370)$ is governed by the non-strange light quark content. Based on
these observations, they have proposed a flavor-mixing scheme to consider the glueball and
$q\bar q$ mixing in the neutral scalar mesons $f_0(1710)$, $f_0(1500)$ and
$f_0(1370)$.\cite{close1} Fits to the measured scalar meson masses and their branching ratios
of strong decays have been performed in several\cite{close1,close2,He} leading to a mixing
matrix of the form
\begin{equation} \label{eq:Close}
 \left(\begin{array}{c } f_0(1370) \\ f_0(1500) \\ f_0(1710)\\ \end{array}\right)=
\left( \begin{array}{c c c} -0.91 & -0.07 & +0.40 \\
                 -0.41 & +0.35 & -0.84 \\
                +0.09 & +0.93 & +0.36 \\
                  \end{array}\right)\left(\begin{array}{c}|N\rangle
                  \\ |S\rangle \\|G\rangle \\ \end{array}\right), \nonumber
\end{equation}
where $|N\rangle$ and $|S\rangle$ denote the quarkonium states $(|u\bar u\rangle + |d\bar
d\rangle)\sqrt{2}$ and $|s\bar s\rangle$, and $|G\rangle$ denotes the pure scalar glueball
state. Thus for Close {\it et al.} the $f_0(1500)$ is composed primarily of a scalar glueball.

Another analysis has been carried out by Cheng {\it et al.}.\cite{CCL} Two lattice results are
employed as the starting point; one is the approximate SU(3) symmetry in the scalar sector
above 1 GeV for the connected insertion part without $q\bar q$ annihilation,\cite{Mathur} and
the other is the scalar glueball mass at 1710 MeV in the quenched
approximation.\cite{Morningstar:1999rf,Chen:2005mg} In the SU(3) symmetry limit, $f_0(1500)$
becomes a pure SU(3) octet and is degenerate with $a_0(1450)$, while $f_0(1370)$ is mainly an
SU(3) singlet with a slight mixing with the scalar glueball which is the primary component of
$f_0(1710)$. These features remain essentially unchanged even when SU(3) breaking is taken
into account. The observed enhancement of $\omega f_0(1710)$ production over $\phi f_0(1710)$
in hadronic J/$\psi$ decays and the copious $f_0(1710)$ production in radiative J/$\psi$
decays lend further support to the prominent glueball nature of $f_0(1710)$. Furthermore,
chiral suppression\cite{Sexton:1995kd,Chanowitz:2005du,chao} is advocated to obtain the
following mixing matrix,
 \begin{equation} \label{eq:wf}
 \left(\begin{array} {c} f_0(1370) \\ f_0(1500) \\ f_0(1710) \\ \end{array}\right)=
\left( \begin{array}{c c c } +0.78 & +0.51 & -0.36 \\
                 -0.54 & +0.84 & +0.03 \\
                +0.32 & +0.18 & +0.93 \\
                  \end{array}\right)\left(\begin{array}{c}|N\rangle \\
 |S\rangle \\ |G\rangle \\ \end{array}\right). \nonumber
 \end{equation}

Therefore for Cheng {\it et al.} it is the $f_0(1710)$ is the
particle composed mostly of a glueball state. Thus a definitive
conclusion on this problem is still lacking.

Finally we would like to mention that
the quarkonium-glueball mixing gives also a strong influence to the
properties of the lowest mass pseudoscalar glueball.\cite{Kochelev:2005tu,Gerasimov:2007sb}

\section{Conclusions} \label{sec:conclusions}

The leitmotif of this review has been that the study of glueballs,
states where the gauge field plays an important dynamical role, is
important to understand the nonperturbative behavior of QCD. This
study requires precise and abundant experimental input, which
despite a lot of effort has not achieved a level of understanding
that allows a unique theoretical interpretation. The lack of clarity
arises from the fact that the theoretical developments are not able
to determine in a well defined manner how the production and decay
properties of glueballs are distinct from those of conventional
mesons. But the future is bright for several reasons. New energy
domains will be studied by BESIII and in the future by
$\overline{\mbox{P}}$ANDA, and therefore the possibility of
producing oddballs, glueballs with exotic quantum numbers will open.
Moreover, at low energy, both Crystal Barrel and in the future GlueX
will produce light mesons with a level of statistics, and precision
for analysis, that will be able to separate exotic from non exotic
behaviors and quark model nonets from particles extraneous to them.
In this process the new theoretical developments will be crucial to
guide and/or interpret the results.

In this review we have discussed three major approaches, lattice QCD, QCD based constituent
models and QCD sum rules, with a short excursion into the AdS/QCD formalism, which at present
leads ultimately to a constituent model type of description.

Due to the lack of observable states, the lattice QCD results have played the role of
experimental data. Lattice QCD is a powerful technique, especially for the determination of
masses. The spectrum of the pure gluodynamics (the pure gauge theory), equivalent to the so
called quenched approximation of lattice QCD, is well established and the calculational errors
are under control. However the various treatments of quark loops (unquenched lattice QCD) and
the effects of the mixing of glueballs with mesons are still a matter of debate and no firm
conclusions can be drawn for the real world. Moreover, important features of the study would
be the determination of the decay properties of the various glueball candidates, and lattice
QCD being a theory described in Euclidean space-time does not possess the capability of
describing asymptotic states and therefore of determining decay properties. However, a full
lattice QCD calculation with unlimited and precise calculational power should produce the
experimental spectrum, if as we strongly believe QCD is the theory of the hadronic
interactions, and moreover, if one can extend lattice QCD into Minkowski space-time one would
be able to determine the decay properties.

Given the lack of experimental and detailed theoretical knowledge,
models turn out to be an interesting laboratory to test ideas of the
various perturbative and nonperturbative mechanisms in the theory.
We have discussed in some detail constituent gluon models for
glueballs from different perspectives. Most of the pioneer work
considered glueballs as strongly bound states of two (or three)
heavy spin-1 gluons. The comparison with lattice QCD shows that the
spectrum obtained by these approaches does not correspond to lattice
QCD. However, if one implements a formalism dealing with only
transverse gluons, the spurious states (induced by the gluonic
longitudinal components) disappear and the hierarchy in the spectrum
of lattice QCD is recovered. These improvements have been developed
up to now only in the two-gluon sectors and the generalization to
negative $C$-parity is still lacking. It remains for the future  to
find out which constituent glueball model confirms the lattice QCD
spectrum for the higher states.

The dynamics in these models is described, in general, by a linear
plus Coulomb potentials. This dynamics, however, does not remove the
degeneracy of the scalar and pseudoscalar glueballs. The candidate
instanton induced interaction, which arises naturally in QCD and
which has very particular properties under spin-parity, is the
natural candidate to do the job. In the quenched approximation, the
instanton induced interaction is attractive in the scalar and
repulsive in the pseudoscalar channels, and equal in magnitude. This
is the precise behavior needed to lift the degeneracy in the
constituent models.

The QCD sum rules offer the possibility to describe mixing and sea quarks effects in glueball
states. The method is rather technical and requires many ingredients (condensates, instantons,
topological charge screening, ...). This complexity has led to many different calculations,
with different results for the lowest lying states. We refer the reader to the literature for
the various attempts to describe the low mass glueball spectrum. In here, we present the
general formalism and some details of a specific calculation, in which the masses are obtained
from pure gluonic currents, however, sea quarks effects are included ({\it via} the quark
condensate).

Glueballs should be produced preferably in gluon-rich processes. We
described briefly the main examples: J/$\psi$ decays, central
production $\bar pp$  annihilation,$\gamma \gamma$ fusion and
photoproduction. The realization of a new generation of experiments,
BESIII, $\overline{\mbox{P}}$ANDA and GlueX, provide hope for new
exciting developments in this field.

The ultra-relativistic heavy ion experiments is another experimental
scenario where one expects that glueballs might play a role. It is
clear that the transition to a different phase of confinement, be it
the Strong Coulomb phase or the Quark Gluon Plasma, precludes that
the confinement properties of the bound systems will change and with
it their physical behavior. There is no consensus on the glueball
properties above the critical temperature and this has led to the
description of several scenarios which can be very interesting from
the point of view of observation. RHIC has produced large amount of
date at high temperature, as will very soon do ALICE. On the other
hand FAIR will probe the high density region. Given these
circumstances it is clear that the aim is to performed realistic
analysis which are are able to disentangle the right from the wrong
ideas.

Finally, the strong expected mixing between glueballs and quark states leads to a broadening
of the possible glueball states which does not simplify their isolation and their theoretical
description. The wishful sharp resonances which would confer the glueball spectra the beauty,
richness and simplicity of the conventional baryonic and mesonic spectra are lacking. We hope
that in the future and at higher energies the situation changes and we are able to have
isolated exotic states. It is, however, important to stress, that in any case gluebals are a
beautiful and unique consequence of QCD.

The study of glueballs is intimately related to the quantitative understanding of confinement
in QCD, since understanding confinement requires an understanding of the soft gluonic field
responsible for binding the hadrons and the structure of the QCD vacuum. We have emphasized
that this has been the goal of the various studies, {\it i.e.} to get a clearer picture of how QCD
behaves at relatively low momentum and how this behavior changes as the temperature or the
density increase. Our review shows that we have learned much but we have yet not achieved the
goal. We foresee a bright future in this respect since a new experimental era of the study of
QCD in many fronts is opening up. To find a description, which is able to describe all the
various phenomena in a unique framework, is our task, and when achieved will imply that we
finally understand QCD. The venture is even more exciting since maybe AdS gravity might come
to our help, a possibility unthinkable a few years back.

\section*{Acknowledgements}
We would like to thank Philippe de Forcrand, Alexander Dorokhov, Hilmar Forkel, Sergo Gerasimov
and Dong-Pil Min for useful discussions. Vincent Mathieu thanks the I.I.S.N.
(Belgium) for financial support. Nikolai Kochelev was partly
supported by Belarus-JINR grant. This work was done Vicente Vento was on a sabbatical from the University of
Valencia at the PH-TH at CERN, whose members he thanks for their hospitality. Vicente Vento was supported
by MECyT-FPA2007 and by MEC-Movilidad PR2007-0048.

\end{document}